\documentclass[fleqn,usenatbib]{mnras}

\usepackage{newtxtext,newtxmath}
\usepackage{hyperref}
\usepackage[T1]{fontenc}
\usepackage[flushleft]{threeparttable}
\DeclareRobustCommand{\VAN}[3]{#2}
\let\VANthebibliography\thebibliography
\def\thebibliography{\DeclareRobustCommand{\VAN}[3]{##3}\VANthebibliography}

\usepackage{graphicx}	% Including figure files
\usepackage{amsmath}	% Advanced maths commands

\title[Peculiar GC Transient Swift J174610.4-290018]{Is the Peculiar Galactic Center Transient Swift J174610.4-290018 a Nova Outburst?}

\author[Z.Q. Hua et al.]{
Ziqian Hua,$^{1,2}$\thanks{E-mail: zqhua@smail.nju.edu.cn}
Zhiyuan Li,$^{1,2,3}$\thanks{E-mail: lizy@nju.edu.cn}
and Zhao Su$^{1,2}$
\\
\\
$^{1}$School of Astronomy and Space Science, Nanjing University, Nanjing 210023, China\\
$^{2}$Key Laboratory of Modern Astronomy and Astrophysics (Nanjing University), Ministry of Education, Nanjing 210023, China\\
$^{3}$Institute of Science and Technology for Deep Space Exploration, Suzhou Campus, Nanjing University, Suzhou 215163, China
}

\date{Accepted XXX. Received YYY; in original form ZZZ}

\pubyear{\the\year{}}

\begin{document}
\label{firstpage}
\pagerange{\pageref{firstpage}--\pageref{lastpage}}
\maketitle

\begin{abstract}
Swift J174610.4–290018 is a peculiar transient X-ray source in the Galactic center. 
First detected by Swift at the onset of an outburst in February 2024, it has since been observed intentionally and serendipitously by multiple X-ray observatories.
To explore its long-term X-ray spectral and temporal behavior, we analyzed archival and recent observations from Chandra, Swift, and NuSTAR spanning from October 2000 to September 2025.
The Chandra data reveal a previously unreported outburst in 2005, followed by an extended quiescent period of $\sim$19 yr with a mean luminosity of $\sim10^{32}~{\rm erg~s^{-1}}$. The 2024 outburst reached a peak 2–8 keV luminosity of ${\rm L_X}\sim10^{35}~{\rm erg~s^{-1}}$ and decayed over $\sim120$ days. 
In both quiescence and outburst, the spectra are well described by a high-temperature ($\sim$10 keV) thermal plasma, featuring prominent emission lines from neutral and highly ionized iron, and tentative chromium lines during the outburst.
The long-term temporal and spectral properties disfavor the accretion disk corona scenario previously proposed based on early XRISM observations. 
Instead, a nova scenario provides a more natural explanation for the observed X-ray flux evolution, spectral characteristics, and possible repeated outbursts, which bear similarity to some known Galactic (recurrent) novae such as RS Oph. 
If confirmed, Swift J174610 would represent the first nova detected in the Galactic center, with important implications for the population of massive white dwarfs and wide binaries near Sgr A*. Continued multi-wavelength follow-up is essential to further elucidate the nature of this remarkable transient.
\end{abstract}


%%%%%%%%%%%%%%%%%%%%%%%%%%%%%%%%%%%%%%%%%%%%%%%%%%

%%%%%%%%%%%%%%%%% BODY OF PAPER %%%%%%%%%%%%%%%%%%

\section{Introduction}
\label{sec:intro}
The Galactic Center (GC) is a unique and dynamic environment characterized by extreme physical conditions, such as high gas densities, strong magnetic fields, and intense radiation fields \citep{Genzel+10}.
Due to its proximity ($\sim$ 8 kpc, \citealp{Do+19,Gravity_collaboration+20}), extensive multi-wavelength observations have revealed the GC's physical structure in detail. 
At its core, the supermassive black hole, commonly known as Sgr A*, is surrounded by a compact and dense nuclear star cluster (NSC) and an extended nuclear star disk (NSD).
These star assemblies are dominated by old, late-type stellar populations, with a substantial fraction residing in binary systems \citep{Feldmeier-Krause+17,Launhardt+02,Sormani+22}.

X-ray observations, particularly those from the {\it Chandra X-ray Observatory} with its superb spatial resolution \citep{Weisskopf+02}, serve as powerful probes of accretion-powered compact binaries.
Numerous past studies primarily using {\it Chandra} observations \citep[e.g.][]{Wang+02,Muno+03,Muno+06,Muno+09,Zhu+18} have revealed that cataclysmic variables (CVs) dominate the X-ray stellar population in the GC while young massive stars (WR/O-type, \citealp{Hua+25}) and low-mass X-ray binaries (LMXBs) are also present.
Long-term, high-cadence monitoring with {\it Chandra}, {\it XMM-Newton}, and {\it Swift} has uncovered a dozen transient X-ray sources \citep{Degenaar+12,Degenaar+15,Ponti+16}.
Most of these transients are believed to be LMXBs whose outbursts reflect substantial changes in the accretion geometry and radiative properties of the system \citep{Bahramian+23}.
With peak X-ray luminosities of $10^{34-36}~{\rm erg~s^{-1}}$ \citep{Wijnands+06}, however, these transient are substantially fainter than typical LMXB outbursts found in other places of the Galaxy \citep{Done+07}, hinting at extraordinary X-ray source populations uniquely present in the GC environment. 
\citet{Zhu+18} showed that the radial surface density distribution of the transients are steeper than those of faint, persistent X-ray sources (typically CVs) in the GC, as expected if the former are dynamically formed binaries.
On the other hand, given the presence of thousands of CVs, it is natural to expect the occurrence of novae in the GC. However, to date there is no unambiguous evidence for nova explosions in the NSC/NSD, partly owing to the strong foreground extinction over the optical to soft X-ray bands. 
Recent hydrodynamic simulations have shown that in the GC environment, nova remnants are expected to evolve rapidly, remain compact, and exhibit only a short-lived X-ray–bright phase, making their detection particularly challenging even with current X-ray observatories \citep{Su+2026}.

On February 22, 2024, the {\it Swift}/XRT instrument detected an excess of point-like X-ray emission during its monitoring campaign of the GC \citep{Reynolds+24}.
The transient, designated Swift J174610.4-290018 (hereafter Swift J174610), is located at RA = $17^{\rm h}46^{\rm m}10.4^{\rm s}$, Dec = $-29^{\circ}00^{\prime}17.6^{\prime\prime}$, with a peak apparent luminosity of $\sim10^{35}~{\rm erg~s^{-1}}$ over 2--10 keV assuming a distance of 8 kpc.
Owing to its proximity ($\sim400^{\prime\prime}$) to Sgr A*, the frequent monitoring of the GC by X-ray observatories including {\it Swift}, {\it NuSTAR}, {\it XRISM}, and {\it Chandra} provides high-cadence observations that help constrain the nature of this transient.

Based on two {\it XRISM/Xtend} observations (with a total exposure of $\sim$170 ks obtained between February 29 and March 2, 2024), \citet{Yoshimoto+25} reported that Swift J174610 exhibited an unusual X-ray spectrum.
Strong emission lines from both helium-like (Fe-XXV, $\sim6.7~{\rm keV}$) and hydrogen-like (Fe-XXVI, $\sim7.0~{\rm keV}$) iron were detected.
The exceptionally high line ratio of $I_{7.0}/I_{6.7} \sim 4$ implies a plasma temperature $T_{\rm line} \sim30 ~{\rm keV}$, which is inconsistent with the much lower bremsstrahlung temperature, $T_{\rm brem}\sim7~{\rm keV}$, determined from the observed continuum over 2--10 keV.
To account for this discrepancy, the authors proposed a scenario in which a neutron-star low-mass X-ray binary (NS-LMXB) is viewed at high inclination. 
In this scenario, the central source with an intrinsic luminosity of $L_{X}\sim10^{37}~{\rm erg~s^{-1}}$ is obscured by its accretion-disk corona (ADC, \citealp{White+82}).
At high inclinations, the direct continuum emission from the neutron star and inner accretion disk is largely blocked, and the observed spectrum is dominated by reprocessed radiation.
The photoionized plasma in the extended corona enhances the hydrogen-like iron line relative to the helium-like line, thereby producing the anomalous $I_{7.0}/I_{6.7}$ ratio.
A short ($\sim$100 sec) X-ray flare detected in a 2004 XMM-Newton observation, originally reported by \citet{Pastor-Marazuela+20} as a candidate Type-I burst, was taken to be further evidence for a NS-LMXB system.
\citet{Stel+25}  used multi-epoch {\it XMM-Newton} observations, along with other X-ray observations, to analyze the X-ray properties of Swift J174610 both during the 2024 outburst and the short flare in 2004. Based on their analysis, these authors also suggested an NS-LMXB/ADC system as the origin of Swift J174610.  
However, these authors did not address the relevance of the transient nature for their ADC scenario.

The wealth of available X-ray data, particularly those from {\it Swift} and {\it Chandra}, allows to track the temporal evolution of Swift J174610, which is key to unveiling its true nature.
We therefore analyzed a large collection of available {\it Swift}, {\it Chandra} and {\it NuSTAR} data to investigate the properties of this transient, paying attention also to its quiescent state. 
The structure of this paper is as follows.
In Section~\ref{sec:data_reduction}, we describe the datasets used and the corresponding reduction procedures.
In Section~\ref{sec:result}, we present the temporal and spectral analysis of Swift J174610.
In Section~\ref{sec:discussion}, we examine potential caveats of the ADC interpretation and propose a recurrent nova origin as an alternative.
Section~\ref{sec:summary} provides a summary of this work.

%%%%%%%%%%%%%%%%%%%%%%%%%%%%%%%%%%%%%%%%%%%%%%%%%%%%%%%%%%%%%%
%%%%%%%%%%%%%%%%%%%%%%%%%%%%%%%%%%%%%%%%%%%%%%%%%%%%%%%%%%%%%%
%%%%%%%%%%%%%%%%%%%%%%%%%%%%%%%%%%%%%%%%%%%%%%%%%%%%%%%%%%%%%%

\section{Data Preparation}
\label{sec:data_reduction}
\begin{figure*}
    \centering
    \includegraphics[width=0.95\linewidth]{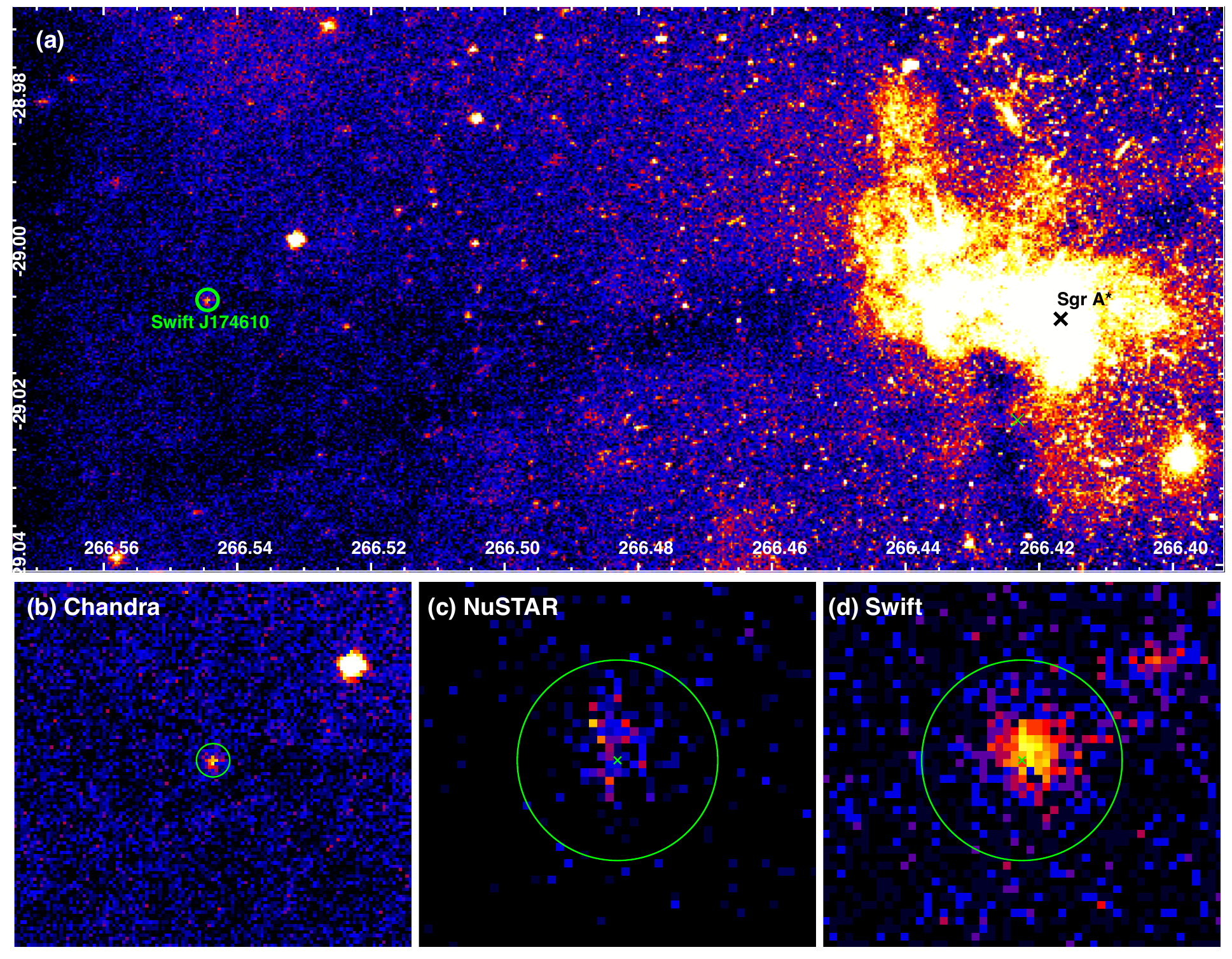}
    \caption{(a) Combined 2–8 keV {\it Chandra} image of the GC, with the horizontal axis aligned with the right ascension direction. 
    The position of Swift J174610 is marked by a green circle of $5^{\prime\prime}$ radius, and Sgr A* is indicated by a black cross.
(b) Zoomed-in {\it Chandra} view of Swift J174610, showing the only likely {\it Chandra} counterpart placed at the center of the panel.
(c) Combined 3–79 keV {\it NuSTAR} image showing the same region as in panel (b). A circle of $30^{\prime\prime}$ radius denotes the source extraction region, and the central cross marks the {\it Chandra}-source position of Swift J174610.
(d) Combined 2–8 keV {\it Swift} image, shown with the same  $30^{\prime\prime}$-radius circle and position marker as in panel (c).
}
    \label{fig:image}
\end{figure*}

Since its initial brightening, Swift J174610 has been frequently monitored by multiple X-ray observatories. 
A search through the HEASARC archive browse\footnote{\url{https://heasarc.gsfc.nasa.gov/cgi-bin/W3Browse/w3browse.pl}} yields 13 {\it Chandra} and 9 {\it NuSTAR} observations, in addition to the nearly daily coverage provided by {\it Swift}.
The corresponding data reduction procedures are described below.
Observations performed with different instruments are combined separately and shown in Fig.~\ref{fig:image}.

\subsection{\it Swift}
\label{sec:swift}
Since its launch in February 2006, {\it Swift} has continuously monitored the GC in $\sim$1 ks exposure with a cadence of every 1–4 days \citep{Degenaar+15}.
This high-cadence program provides systematic coverage of the 2024 outburst of Swift J174610 from Feb 22 to 6 June, when the source had become too faint to be detected (another possible outburst occurred before {\it Swift} was operational; see Section~\ref{sec:burst2005}).
We analyzed all available {\it Swift}/XRT observations since February 2023 up to September 2025, which help to determine the quiescent level before the 2024 outburst and to trace its post-outburst evolution (see a long-term {\it Swift} light curve in Fig.~\ref{fig:swift_lc}).
In total, 582 observations with a summed exposure of $\sim 489$ ks (80 observations with $\sim 70$ ks exposure covering the outburst) were uniformly reprocessed using the standard {\it xrtpipeline} task in HEASoft v6.35.
Source and background spectra were extracted with {\it xselect}, with the source extracted from a circular region enclosing the 90\% enclosed counts radius (ECR), and the background from an annular region with 2–4 times this radius, estimated from the point spread function (PSF) as a function of off-axis angle and photon energy following \citet{Moretti+05}.
The net count rates were converted to unabsorbed fluxes using the best-fit spectral model.
The source centroid was fixed to the position determined from the {\it Chandra} data (Section~\ref{sec:chandra_data}).
For spectral analysis, we used only the exposures obtained during February--April 2024 (totaling $\sim13$ ks), when the transient maintained a high 2–8 keV flux (Fig.~\ref{fig:lc}).
The spectral analysis was performed with XSPEC v12.15.0 \citep{Arnaud+96}.
Using the same extraction regions, we constructed the 2–8 keV X-ray light curve from all available Swift/XRT observations. 
For each observation, source and background counts were measured and the net count rates, along with their uncertainties, were estimated using the {\it aprates} tool. 
To improve the signal-to-noise ratio, every five consecutive observations were grouped together.

\subsection{\it Chandra}
\label{sec:chandra_data}
The GC has been the target of extensive {\it Chandra} observations since 1999, chiefly to investigate the SMBH's variability, its surrounding diverse X-ray stellar populations, and molecular clouds irradiated by Sgr A*.
Swift J174610 falls within the coverage of most of these observations, which provide a total exposure exceeding 2.5 Ms over a 25-year timespan.
Such a dataset enables a detailed study of the transient’s long-term variability, revealing its properties in quiescence and offering valuable insights into its nature.
Meanwhile, {\it Chandra}'s superb angular resolution delivers the most precise localization of Swift J174610, aiding in future identification of possible multi-wavelength counterparts.

A total of 113 observations covering the source position are available.
The data were reduced using CIAO v4.14 and CALDB v4.9.2.
Common sources were identified among individual observations and aligned to the reference frame defined by the longest exposure, ObsID 3392, for relative astrometric calibration.
For each observation, we generated count maps, exposure maps, and 90\%-ECR maps in the 2–8 keV band, assuming a bremsstrahlung spectral model with a plasma temperature of 10 keV and a Galactic column density of $N_{\rm H} = 10^{23}~{\rm cm^{-2}}$.
The source position was initially identified with the CIAO task {\it wavdetect} and subsequently refined through a maximum-likelihood fitting procedure \citep{Boese+01}. 
The best-fit centroid, RA=${\rm 17^{h}46^{m}10.67^{s}}$ and Dec=${\rm -29^{\circ}00^{\prime}19.44^{\prime\prime}}$ is in excellent agreement with source No.3596 in the NSC X-ray source catalog of \citet{Zhu+18}.
This source is clearly the only possible counterpart to the {\it NuSTAR} and {\it Swift} detections (Fig.~\ref{fig:image}).
Source photons were extracted within the 90\% ECR, while background photons were taken from an annular region with radii of 2–4 times the 90\% ECR.
The 2--8 keV fluxes of Swift J174610 and their corresponding 1-$\sigma$ uncertainties were computed with CIAO tool {\it aprates} and are shown in Fig.~\ref{fig:lc} (upper limits are indicated for non-detections).
The source spectra from individual observations were extracted using the CIAO tool {\it specextract}.

\subsection{\it NuSTAR}
Between April 4 -- 12, 2024, {\it NuSTAR} conducted 9 observations of the Sgr A* region, which also covered Swift J174610 with a total exposure of $\sim 150$ ks.
Thanks to its sensitivity in the hard X-ray band (3–79 keV), {\it NuSTAR} provides crucial information on the source’s overall spectral characteristics, helping to constrain the plasma temperature.
Data reduction was performed using the HEASoft v6.35 task {\it nupipeline}.
A consistent 30$''$-radius extraction region was applied to all observations, while the background was extracted from an annular region with inner and outer radii of 60$''$ and 120$''$, respectively.
The NuSTAR light curve in the 2–8 keV band was extracted from this region and the count rates were converted to fluxes using the best-fit spectral model.
Spectra from individual observations were generated with the HEASoft tool {\it nuproducts} and combined separately for the two focal-plane modules, FPMA and FPMB, but fitted simultaneously.

%%%%%%%%%%%%%%%%%%%%%%%%%%%%%%%%%%%%%%%%%%%%%%%%%%%%%%%%%%%%%%
%%%%%%%%%%%%%%%%%%%%%%%%%%%%%%%%%%%%%%%%%%%%%%%%%%%%%%%%%%%%%%
%%%%%%%%%%%%%%%%%%%%%%%%%%%%%%%%%%%%%%%%%%%%%%%%%%%%%%%%%%%%%%

\begin{figure*}
    \centering
    \includegraphics[width=0.95\linewidth]{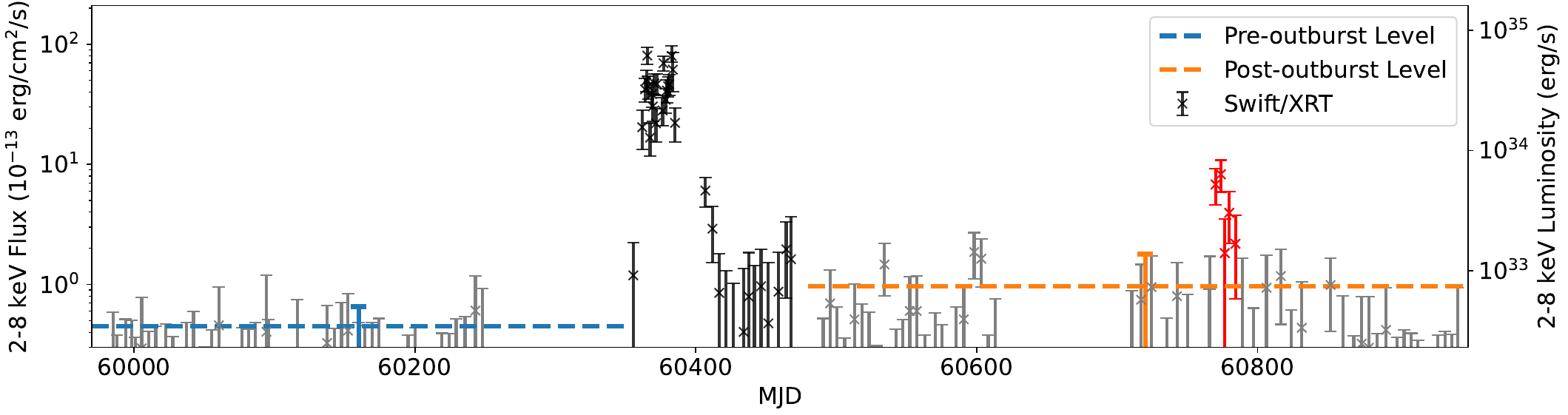}
    \caption{The 2-8 keV {\it Swift}/XRT light curve of Swift J174610 between 2023 and 2025. Every five consecutive observations are grouped to enhance the $S/N$. Observed count rates have been converted to 2--8 keV unabsorbed energy fluxes using the best-fit spectral models. The blue and red dashed lines represent the mean flux level of the pre-outburst epochs and post-outburst epochs (excluding the short flaring episode denoted in red), respectively.}
    \label{fig:swift_lc}
\end{figure*}

\section{Analysis and Results}
\label{sec:result}
\subsection{Long-term Variability}
\label{sec:long-term var}
\begin{figure*}
    \centering
    \includegraphics[width=0.95\linewidth]{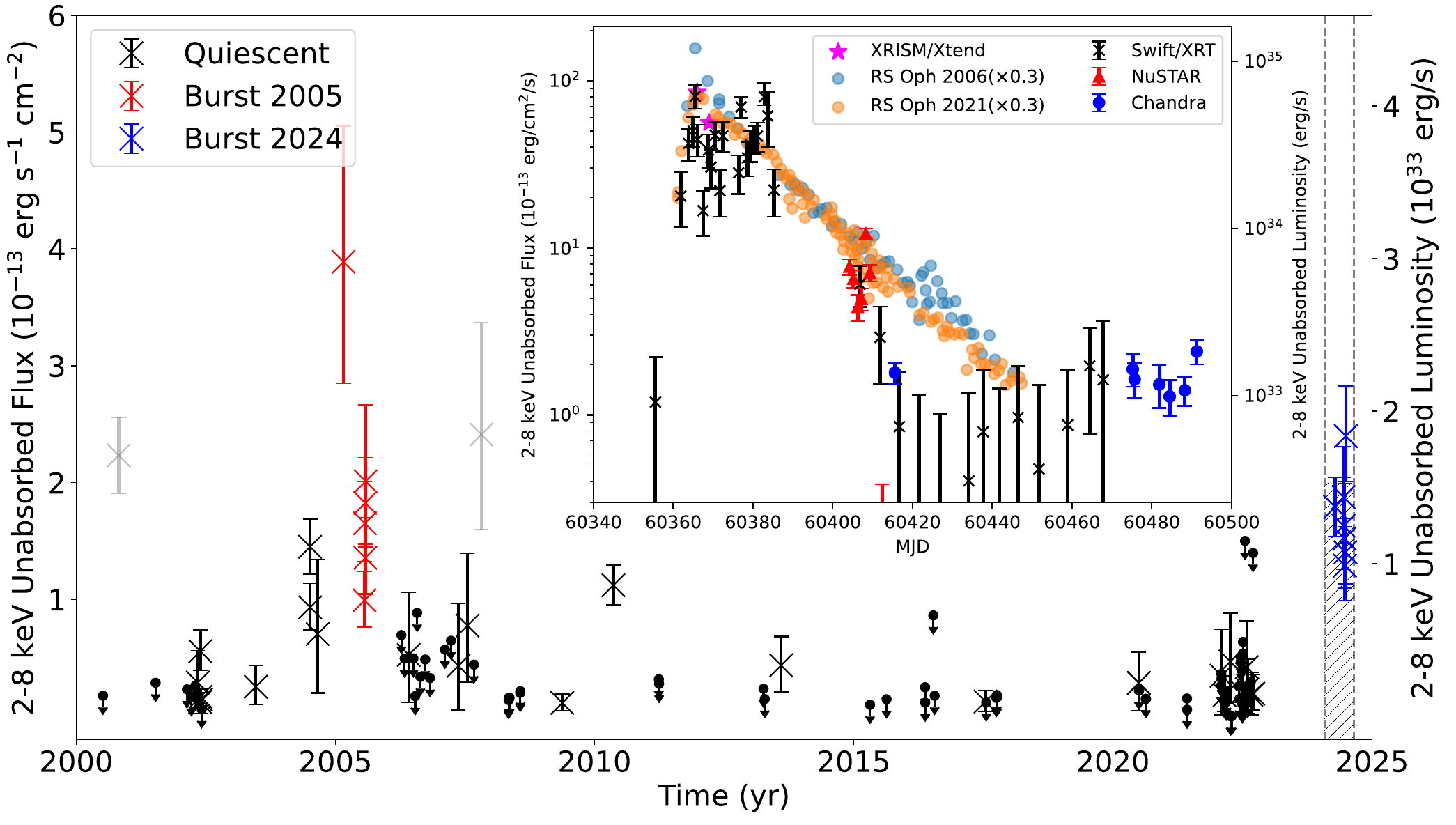}
    \caption{{\it Main panel:} The 2--8 keV X-ray light curve of Swift J174610, spanning from 2000 to 2025 as observed by {\it Chandra}/ACIS. 
    The unabosrbed flux and its associated $1 \sigma$ error (denoted by crosses), or its 3$\sigma$ upper limits in the case of a non-detection (denoted by downward arrows), are derived using CIAO tool {\it aprates} for each observation, taking into account the Poisson statistics in the low-count regime.  
    The observations utilized for spectral extraction of the 2005 outburst, 2024 outburst, and the quiescent state are denoted by blue, orange, and black markers, respectively, while the remaining observations are shown in grey. The shaded region encloses the timespan of the insert, which presents a cross-observatory view of the 2--8 keV  flux evolution of the 2024 outburst. 
    Observed count rates from {\it Swift} (black crosses), {\it NuSTAR} (red triangles), and {\it Chandra} (blue circles) and {\it XRISM} (magenta stars, as reported by \citealp{Yoshimoto+25}) have been converted to 2--8 keV energy fluxes using their respective best-fit spectral models, and further to the unabsorbed luminosity by assuming a distance of 8 kpc.
    To enhance the signal-to-noise ratio during the source's fading phase, we grouped every five consecutive {\it Swift} observations taken since April 2024. 
    For comparison, the 1--10 keV light curves of two historical outbursts (2006 and 2021; \citealp{Page+22}) from the recurrent nova RS Oph are plotted by light blue and orange symbols.
    The luminosities have been renormalized to a distance of 8 kpc and converted to the 2–8 keV energy band. 
    A rescaling factor of 0.3 is applied for better comparison.
    The peak times are aligned with the flux maximum of Swift J174610 (MJD 60365.5).
}
    \label{fig:lc}
\end{figure*}

Given the abundant observations of Swift J174610, we first utilized the {\it Chandra} archival data to investigate its pre-2024 long-term variability.
As shown in Fig.~\ref{fig:lc}, Swift J174610 was detected in a number of {\it Chandra} observations as early as 2001, with a 2--8 keV unabsorbed flux of $\gtrsim 3\times10^{-14}\rm~erg~cm^{-2}~s^{-1}$, while in the remaining observations the source was too faint to be detected. 
Interestingly, in a group of observations taken between February -- July 2005, Swift J174610 exhibited an elevated flux level (with the highest flux $\sim 4\times10^{-13}\rm~erg~cm^{-2}~s^{-1}$ recorded on February 27, 2005), comparable to that of the 2024 outburst in its intermediate-to-late phases (see insert of Fig.~\ref{fig:lc}). 
This indicates that Swift J174610 might have experienced an earlier outburst.
Unfortunately, this tentative 2005 outburst was only sparsely covered by {\it Chandra} and predated the {\it Swift} monitoring campaign. It is thus possible that even higher fluxes existed shortly before or after February 27. 
Outside the two outburst episodes, Swift J174610 were generally faint, and no evidence of gradual long-term evolution is found.

The long-term monitoring by Swift/XRT further constrains the source behavior immediately before and after the 2024 outburst.
A total of 267 and 235 Swift observations obtained in 2023 and 2025, respectively, were used to trace the pre-outburst and post-outburst evolution.
As shown in Fig.~\ref{fig:swift_lc}, the source remained at a low flux level throughout 2023 (MJD 59981–60248), with an average unabsorbed 2–8 keV luminosity of $\sim3\times10^{32}~{\rm erg~s^{-1}}$.
In 2025, a very brief flaring episode was observed (red symbols in  Fig.~\ref{fig:swift_lc}), reaching a luminosity of $\sim7\times10^{33}~{\rm erg~s^{-1}}$ on MJD 60773 and lasting for about 10 days, as also reported by \citet{Stel+25}. 
While the peak flux of this flaring is comparable to that of the plausible 2005 outburst revealed by Chandra, its much shorter duration argues against a distinct outburst of the same nature.
Excluding this brief flaring phase, the average unabsorbed 2–8 keV luminosity after the 2024 outburst (MJD 60490–60942) is $\sim8\times10^{32}~{\rm erg~s^{-1}}$. 
The lack of substantial flux enhancement in early 2023 further rules out an outburst or flaring recurrence on a timescale of $\sim$1 yr.

We have also searched for potential periodic modulations using both the Gregory–Loredo and Lomb–Scargle algorithms to the X-ray light curves obtained by the different instruments, but no significant periodicity was detected, lending no support to the claim by \citet{Yoshimoto+25} for a periodic signal of 1537 s in the {\it XRISM/Xtend} observations.
Nor did we find any short timescale ($\lesssim$100 sec) flare similar to that seen in the 2004 XMM-Newton observation \citep{Pastor-Marazuela+20} in any of the {\it Chandra} observations.
In the following sections, we separately examine the properties of Swift J174610 during its quiescent state and during its two outbursts.

\subsection{Quiescent state}
\label{sec:qstate}
To technically define a quiescent state in clear distinction from the two outbursts, we adopted a threshold of $1\times10^{-13}~{\rm erg~s^{-1}~cm^{-2}}$ (Figure~\ref{fig:lc}), which corresponds to an unabsorbed luminosity of $\sim 7\times10^{32}~{\rm erg~s^{-1}}$ at a distance of 8 kpc given the best spectral model (see below).
This threshold is also consistent with the 2023 and 2025 Swift/XRT monitoring.
Since Swift observations have insufficient photons for spectral extraction, the quiescent-state spectrum was formed by co-adding the relevant individual {\it Chandra} observations and is displayed in Fig.~\ref{fig:spec}(a), which clearly exhibits emission line features between $\sim$ 6--7 keV, on top of a continuum heavily absorbed below $\sim$ 2 keV.

We began our spectral analysis with an absorbed optically-thin thermal plasma model ({\tt TBabs*apec} in XSPEC), assuming solar abundances.
Such a model serves for a dual purpose, which provides a basic characterization of the quiescent-state spectrum and enables consistent flux conversions from the observed count rate across different observations and different instruments. 
The quiescent-state spectrum is well described by the model, with a best-fit plasma temperature of $9.7^{+7.3}_{-4.5}$ keV and an unabsorbed 2--8 keV luminosity of $\sim 1.6\times10^{32}~{\rm erg~s^{-1}}$.
To further quantify the emission line features, we applied an alternative phenomenological model, {\tt TBabs*(bremss+4*gauss)}, in which a maximum of four Gaussian lines were included in addition to a bremsstrahlung continuum, the centroids of which were fixed at 5.65, 6.4, 6.7, and 7.0 keV. 
These correspond to helium-like Cr (marginally detected in the quiescent-state but prominent in the outburst spectra; see Fig.~\ref{fig:spec}), neutral Fe K$\alpha$, helium-like Fe K$\alpha$, and hydrogen-like Fe Ly$\alpha$, respectively. 
No intrinsic line broadening was introduced.
The thus measured line ratios are $I_{7.0}/I_{6.7}=0.65^{+0.24}_{-0.21}$ and $I_{6.4}/I_{6.7}=0.46^{+0.16}_{-0.15}$.
The heavy foreground absorption constrained by the fitted ${\rm N_H}\sim2\times10^{23}~{\rm cm^{-2}}$, combined with the absence of significant redshift in the lines, safely places Swift J174610 at or near the GC.

\subsection{Outburst state}
\label{sec:burst}

\subsubsection{2005 Outburst}
\label{sec:burst2005}
Because of the sparse temporal sampling and the lack of complementary observations from other instruments, it is difficult to reconstruct the detailed profile of the 2005 outburst.
Nevertheless, the {\it Chandra} data, with a total exposure of $\sim250$ ks, provide a high-quality combined spectrum (Fig.~\ref{fig:spec}b).
Compared with the quiescent-state spectrum, a prominent feature emerges at $\sim$5.6 keV.
This feature is likely the Cr XXIII K$\alpha$ transition, whose rest-frame energy is 5.65 keV.
This tentative identification is in accord with the $\sim$5.9 keV excess in the {\it XRISM} spectra reported by \cite{Yoshimoto+25} during the 2024 outburst, which was attributed to Cr XXIV Ly$\alpha$.
In addition, significant Fe lines are present, with derived line ratios of $I_{7.0}/I_{6.7}=0.60^{+0.16}_{-0.12}$ and $I_{6.4}/I_{6.7}=1.10^{+0.33}_{-0.21}$.
The best-fit plasma temperature ($8.5^{+5.1}_{-3.6}$ keV) is comparable to that in the quiescent state.

\subsubsection{2024 Outburst}
\label{sec:burst2024}
As shown in Fig.~\ref{fig:lc}, {\it Swift} first detected the 2024 outburst on February 22 and monitored the source until March 15.
The monitoring resumed on April 4 and continued until June 6.
{\it NuSTAR} observed Swift J174610 in April, while most {\it Chandra} observations were performed in June.
Overall, the source maintained at a high, but fluctuating, flux level ($\sim 2-8 \times 10^{-12}\rm~erg~cm^{-2}~s^{-1}$) in the pre-April {\it Swift} observations.
There may exist a possible secondary maximum during this phase (see Fig.~\ref{fig:nova_lc}) which will be discussed in Section~\ref{sec:recurrent_nova} in detail.
Then Swift J174610 faded by about an order of magnitude in the {\it NuSTAR} epochs, and further declined to a flux level of a few $10^{-14}\rm~erg~cm^{-2}~s^{-1}$ by May 2024 (MJD 60450, i.e. $\sim$80 days since the peak).
There also appears to be a re-brightening or flattening episode beginning June 2024 (MJD 60460), dictated by the latest {\it Swift} and {\it Chandra} observations.

\begin{figure*}
    \centering
    \includegraphics[width=0.495\linewidth]{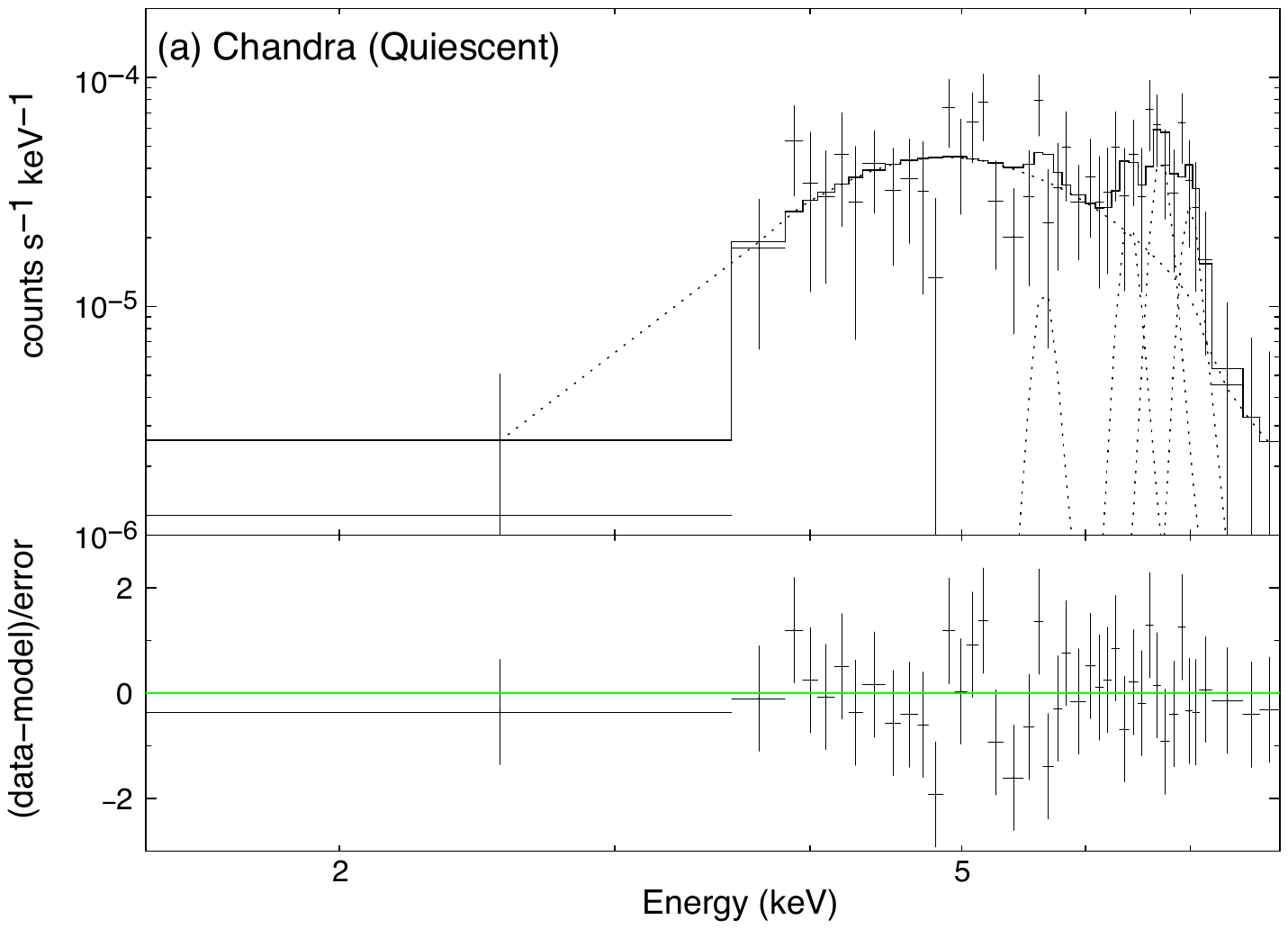}
    \includegraphics[width=0.495\linewidth]{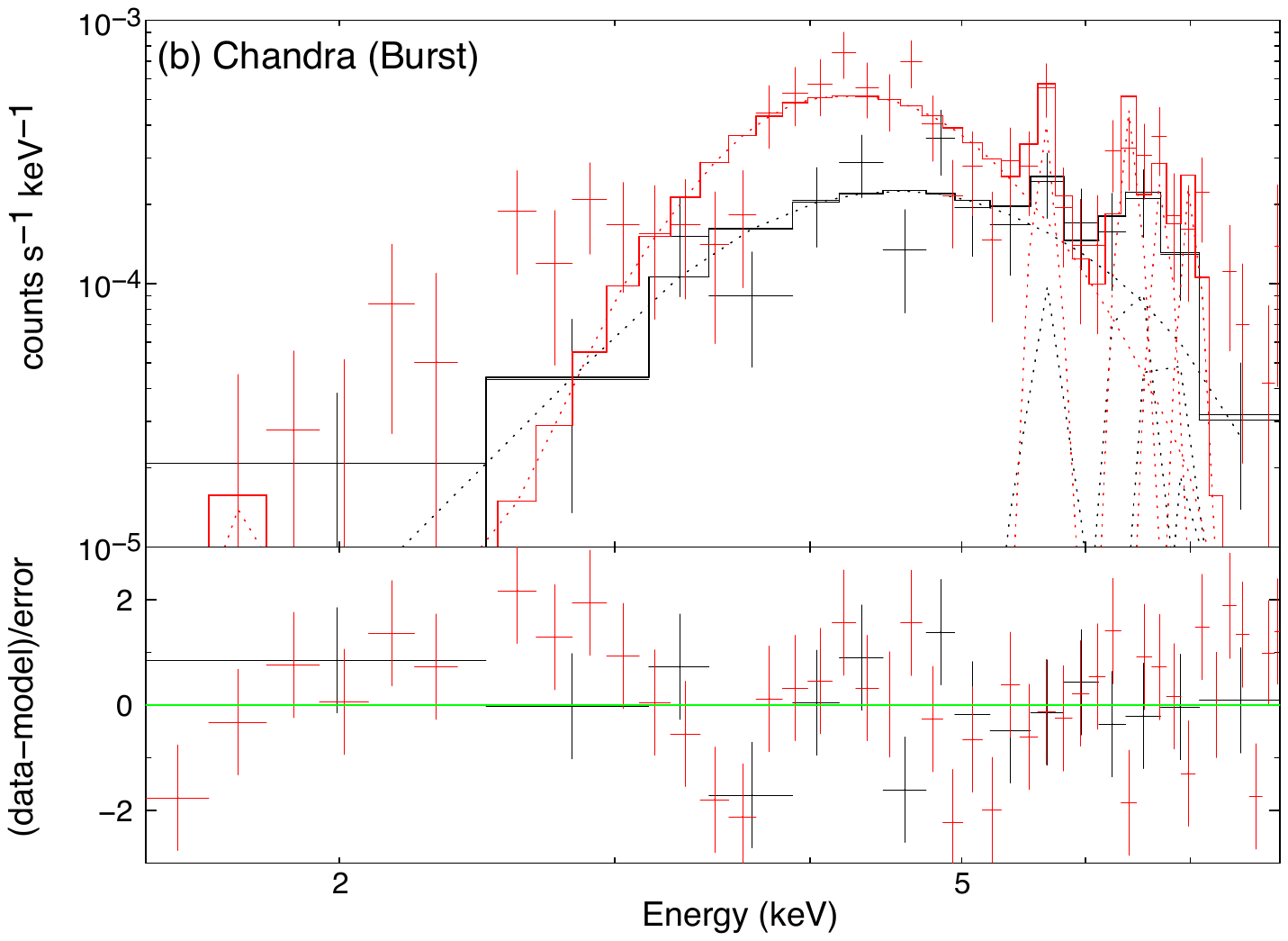}
    \includegraphics[width=0.495\linewidth]{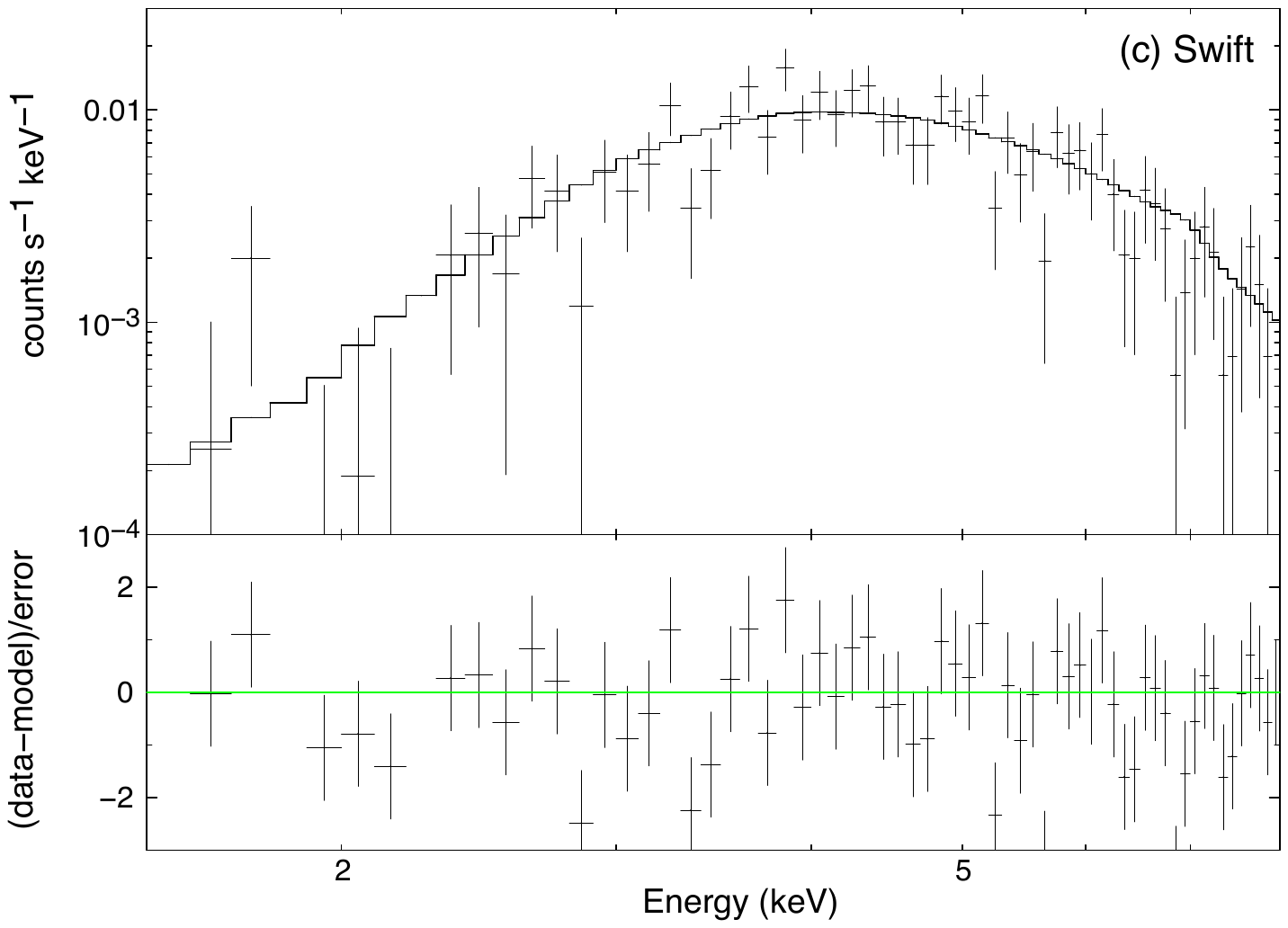}
    \includegraphics[width=0.495\linewidth]{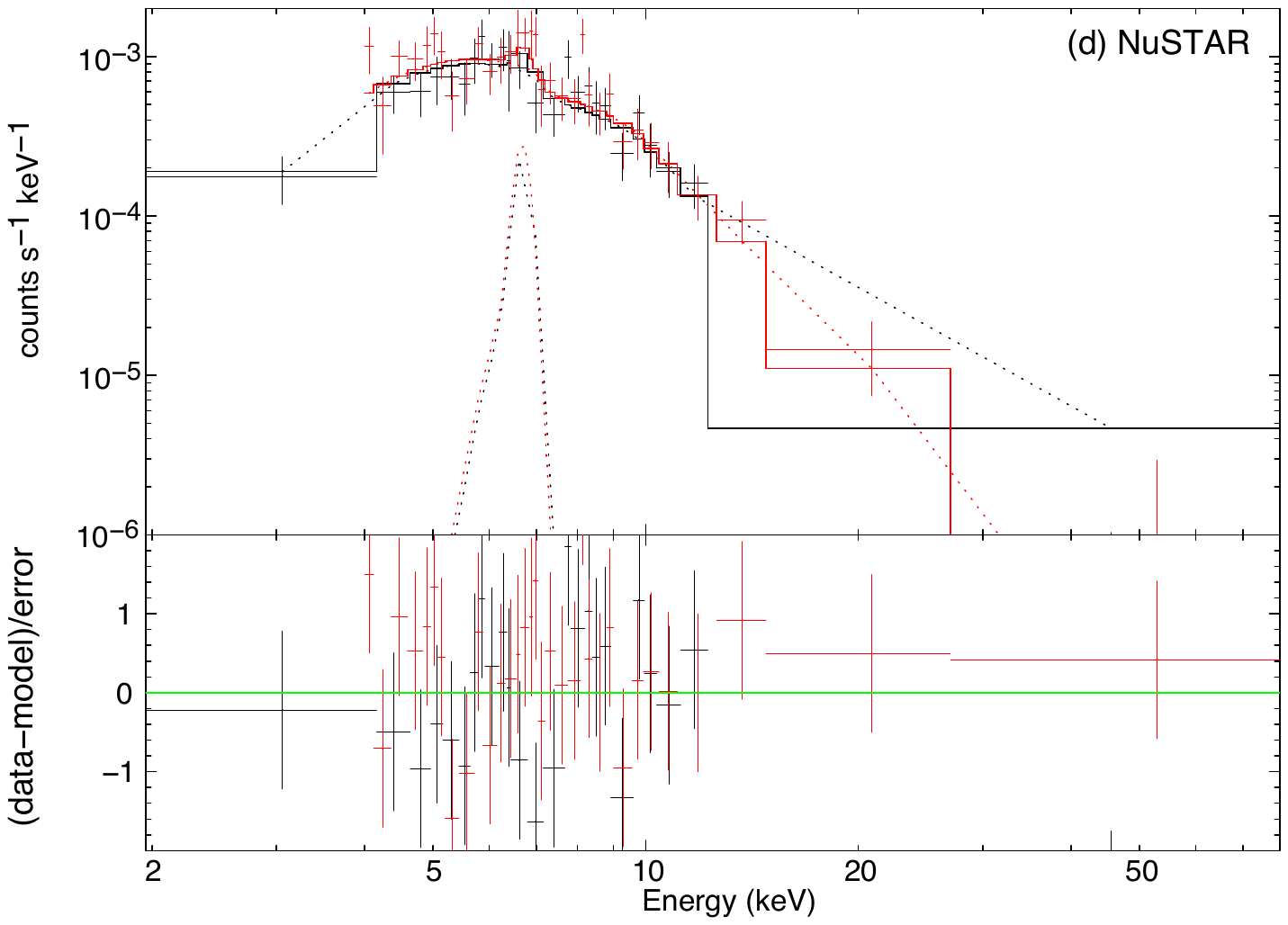}
    \caption{Background-subtracted spectra of Swift J174610 at different flux levels.
    {\bf (a)} The quiescent state observed by {\it Chandra}. 
    {\bf (b)} The two outbursts observed by {\it Chandra}, black for 2005 and red for 2024.
    {\bf (c)} The 2024 outburst observed by {\it Swift} before April.
    {\bf (d)} The 2024 outburst observed by {\it NuSTAR}. The combined spectra of FPMA and FPMB are shown in red and black, respectively.
   The spectra are adaptively binned to ensure a minimum of 10 counts per spectral bin and a S/N $\geq$ 3.
	The best-fit model in XSPEC, {\tt TBABS*(BREMSS+4*GAUSS)} is overlaid as solid curves. A maximum of four Gaussian lines with fixed centroids at 5.65, 6.4, 6.7, and 7.0 keV, are shown by the dotted curves.
	The bottom panels display the residuals, with 1$\sigma$ error bars.}
    \label{fig:spec}
\end{figure*}

No significant line features are found in the combined pre-April {\it Swift} spectrum (Fig.~\ref{fig:spec}c), likely due to the moderate sensitivity despite an intrinsically high flux level. The XRISM spectra taken at around the same time exhibited strong emission lines of Fe and Cr \citep{Yoshimoto+25}.
Nevertheless, the absorbed APEC model gives a high plasma temperature of $\sim$20 keV, consistent with the XRISM result.
Similarly, the combined {\it NuSTAR} spectrum shows no clear evidence for emission lines except for the 6.7 keV line (Fig.~\ref{fig:spec}d), likely due to its limited energy resolution.
The fitted plasma temperature drops to $\sim$7 keV in the {\it NuSTAR} epochs.
On the other hand, the combined {\it Chandra} spectrum closely resembles those obtained during the 2005 outburst (Fig.~\ref{fig:spec}b), showing pronounced excesses at 5.6 and 6.4 keV.
The measured line ratios are $I_{7.0}/I_{6.7}=1.15^{+0.17}_{-0.23}$ and $I_{6.4}/I_{6.7}=1.58^{+0.33}_{-0.27}$.

All spectral fit results are summarized in Table~\ref{tab:fitting_result}.
The best-fit models of absorbed bremsstrahlung plus Gaussian lines are plotted in Fig.~\ref{fig:spec}.

\begin{table*}
\caption{Spectral Fit Results}
\label{tab:fitting_result}
\begin{threeparttable}
\centering
\begin{tabular}{lccccccc}
\hline\hline
 & $N_{\rm H}$ & $kT$ & EW$_{6.7}^{(a)}$ & $I_{6.4}/I_{6.7}$ & $I_{7.0}/I_{6.7}$ & $L_{\rm 2-8\,keV}^{(b)}$ & $\chi^2/{\rm d.o.f.}$ \\
 & $(10^{22}\,{\rm cm^{-2}})$ & (keV) & (eV) & & & (erg\,s$^{-1}$) & \\
\hline
{\it Chandra} (Quiescent)  & $26.4^{+15.1}_{-10.0}$ & $9.7^{+7.3}_{-4.5}$ & $317^{+33}_{-25}$  & $0.46^{+0.16}_{-0.15}$ & $0.65^{+0.24}_{-0.21}$ & $1.6^{+0.1}_{-0.1}\times10^{32}$ & 118.23/204 \\
{\it Chandra} (2005 Outburst) & $24.7^{+11.5}_{-7.0}$ & $8.5^{+5.1}_{-3.6}$ & $240^{+17}_{-14}$ & $1.10^{+0.33}_{-0.21}$ & $0.60^{+0.16}_{-0.12}$ & $1.8^{+0.2}_{-0.1}\times10^{33}$ & 115.52/156 \\
{\it Chandra} (2024 Outburst) & $16.5^{+4.3}_{-3.7}$ & $8.2^{+7.1}_{-4.3}$ & $204^{+14}_{-13}$ & $1.58^{+0.33}_{-0.27}$ & $1.15^{+0.17}_{-0.23}$ & $1.2^{+0.1}_{-0.1}\times10^{33}$ & 197.31/204 \\
{\it Swift} (Feb--Mar 2024) & $11.2^{+1.8}_{-1.6}$ & $21.3^{+17.7}_{-12.6}$ & $<35^{(c)}$ & -- & -- & $8.5^{+0.7}_{-0.6}\times10^{34}$ & 642.08/649 \\
{\it NuSTAR} (Apr 2024) & $15.6^{+2.3}_{-1.9}$ & $7.1^{+3.1}_{-2.4}$ & $150^{+39}_{-37}$ & -- & -- & $5.1^{+0.5}_{-0.4}\times10^{33}$ & 992.12/1004 \\
\hline
\end{tabular}
\begin{tablenotes}
	\small
	\item
	Notes:
Fitted or derived parameters based on the {\tt TBabs*apec} and {\tt TBabs*(bremss+4*gauss)} models.  
Reported errors are at the 90\% confidence level.  
See Sect.~X for details on data sets and spectral modelling.  
$^{(a)}$ Equivalent width of the 6.7 keV line.  
$^{(b)}$ Unabsorbed 2--8 keV luminosity.  
$^{(c)}$ 3$\sigma$ upper limit.
	\end{tablenotes} 
\end{threeparttable}
\end{table*}

%%%%%%%%%%%%%%%%%%%%%%%%%%%%%%%%%%%%%%%%%%%%%%%%%%%%%%%%%%%%%%
%%%%%%%%%%%%%%%%%%%%%%%%%%%%%%%%%%%%%%%%%%%%%%%%%%%%%%%%%%%%%%
%%%%%%%%%%%%%%%%%%%%%%%%%%%%%%%%%%%%%%%%%%%%%%%%%%%%%%%%%%%%%%

\section{Discussion: Origin of Swift J174610}
\label{sec:discussion}
The temporal and spectral properties of Swift J174610 together make it a highly unusual source among the thousands of X-ray sources ever detected in the GC.
While the combination of a $\sim$10 keV bremsstrahlung and prominent Fe lines are characteristic of bright CVs found in the NSC \citep{Zhu+18,Xu+19}, the observed peak 2--8 luminosity of Swift J174610 reaching $\sim 10^{35}\rm~erg~s^{-1}$ is rather unlikely achievable by normal CVs or classical novae, which have typical hard X-ray luminosities of $\lesssim10^{33}\rm{~erg~s^{-1}}$ and $\sim10^{34}\rm{~erg~s^{-1}}$, respectively \citep{Mukai+17}. 
Similarly, while colliding wind binaries involving two massive stars generally exhibit a thermal X-ray spectrum, both the high X-ray luminosity and high plasma temperature are hardly seen in such systems \citep{Hua+25}. 
This invites the consideration of more exotic origins for Swift J174610.
Below we first point out some difficulties with the ADC scenario, and then propose a nova scenario for the observed X-ray behavior of Swift J174610. 

\subsection{The accretion disk corona scenario}
Based on {\it XRISM/Xtend} observations, \citet{Yoshimoto+25} proposed that Swift J174610 is an NS-LMXB viewed at a high inclination angle.
In this scenario, the system hosts a hot, extended ADC \citep{White+82} that scatters and reprocesses a large portion of the immediate X-ray emission from the accretor.
Several Galactic X-ray binaries have been identified as ADC sources, such as X1822-371 \citep{Iaria+13}, 2S 0921-630 \citep{Yoneyama+23} and 4U 2129+47 \citep{Nowak+02}.
Swift J174610 shares several observational characteristics with these objects, including its relatively low observed luminosity ($\sim10^{35}~\rm{erg~s^{-1}}$) and the presence of prominent iron lines.
\citet{Pastor-Marazuela+20} also reported a candidate Type-I burst event from an XMM-Newton observation in 2004, which, if true, might be supportive of an NS-LMXB origin.

The ADC interpretation requires that a substantial fraction of the observed X-ray emission, across different luminosity states, be processed by scattering in an extended accretion disk corona.
Such a corona can only be sustained at relatively high accretion rates, corresponding to a neutron star in a high/soft state with an intrinsic luminosity of ${\rm L_X}\sim10^{37}{\rm~erg~s^{-1}}$.
At these luminosities, intense irradiation from the neutron star and the inner accretion disk or boundary layer can heat and evaporate the upper layers of the disk, producing an extended, ionized atmosphere that forms the ADC.
However, {\it Chandra} observations of Swift J174610 in quiescence reveal a much lower observed luminosity of ${\rm L_X}\sim10^{32}{\rm~erg~s^{-1}}$, about three orders of magnitude below the observed peak value.
Even allowing for heavy obscuration and scattering in the ADC, a simple scaling would imply an intrinsic luminosity of only 
 ${\rm L_X}\sim10^{34}{\rm~erg~s^{-1}}$, which is well below the typical luminosity of even low/hard–state LMXBs (${\rm L_X}\sim10^{36}{\rm~erg~s^{-1}}$; \citealp{Done+07}).
At such suppressed accretion rates and irradiation levels, an extended ADC is not expected to survive, rendering the ADC interpretation of the quiescent-state spectrum, including the prominent iron lines observed by {\it Chandra}, physically implausible.

In addition, observed at high inclination, ADC sources usually display orbital dips and eclipses in their light curves due to obscuration by the inflated disk or the companion star.
The lack of periodic signals detected from Swift J174610 further disfavors the ADC interpretation.
Instead, the source exhibited rapid variability a few days after the first {\it Swift} detection (Fig.~\ref{fig:lc}).
We considered but ruled out the possibility that foreground dust scattering had caused this rapid variability (see quantitative discussions in Appendix~\ref{sec:dust}), which is most likely intrinsic.
In canonical ADC systems, mass transfer is relatively steady, and any rapid changes in the boundary layer emission are smoothed out by reprocessing in the corona \citep{Church+04}.

Moreover, before Swift J174610, about a dozen transient X-ray sources were discovered by {\it Swift} in the NSC/NSD \citep{Degenaar+12,Degenaar+15}. While these transients are generally thought to be LMXBs for their peak luminosities of $10^{35-36}\rm~erg~s^{-1}$, none of them is known to exhibit significant Fe lines in the X-ray spectrum despite sufficient data sensitivity, either during outburst or in quiescence. 
These contrasts with Swift J174610 suggest that it is not an LMXB and strongly disfavor the ADC scenario.

A further problem arises when considering the flare event reported by \citet{Pastor-Marazuela+20}.
\citet{Stel+25} claimed that all the observed flux originates from the emission scattered by the ADC.
If the flare were indeed strongly affected by the ADC scattering, then the intrinsic peak luminosity could be significantly higher than the observed value and fall within the typical range of intermediate-duration Type-I bursts ($10^{38}$--$10^{39}~\rm{erg~s^{-1}}$, \citealp{Alizai+23}).
However, scattering in the extended corona would inevitably smear and distort the intrinsic burst profile, making it difficult to preserve the sharp rise and decay characteristic of thermonuclear Type-I bursts.
More importantly, the high accretion rate required to maintain an extended ADC is in direct contrast to the physical conditions inferred for intermediate-duration bursts \citep{Alizai+23}.
They are thought to originate from thermonuclear ignition in a thick helium layer at relatively low accretion rates, allowing sufficient fuel to accumulate over long timescales.
Furthermore, our inspection of all relevant {\it Chandra} observations revealed no additional flare events over a total exposure of 2.5 Ms in the past 25 yrs.
This is inconsistent with the recurrent nature of Type-I bursts expected under the high accretion rates implied by the ADC scenario.

On the other hand, if one assumes that the flare emission is not significantly affected by the ADC, its unusually long duration becomes difficult to explain.
While long Type-I bursts do exist in some Galactic NS-LMXBs, their peak bolometric luminosities are still generally found to be close to the Eddington luminosity of an NS \citep{Alizai+23}, i.e., $L_{\rm bol} \gtrsim 10^{38}~\rm{erg~s^{-1}}$, which is about an order of magnitude higher than the value inferred for the candidate burst \citep{Stel+25}.
For the above reasons, we suggest that the 2004 flare is unlikely to be a thermonuclear Type-I burst.

\subsection{A nova outburst in the GC?}
\label{sec:recurrent_nova}
According to \citet{Mukai+08}, classical and recurrent novae (RNe) may account for a subset of the faint ($10^{34}$--$10^{35}~{\rm erg~s^{-1}}$) X-ray transients detected in the GC.
A \textbf{nova outburst (possibly with a symbiotic campanion)} as the likely origin of the 2024 outburst of Swift J174610 is a natural proposition, in view of the following observational facts: 
(i) the relatively high X-ray luminosity, high plasma temperature and prominent Fe lines observed during the 2024 outburst (Section~\ref{sec:burst}) are shared properties of some of the best-studied symbiotic novae\citep{Sokoloski+06,Islam+24}; 
(ii) a plausible outburst in 2005, predated the 2024 one by about 19 yrs; 
(iii) the quiescent-state spectrum shares similarities (luminosity, temperature and Fe lines; Section~\ref{sec:qstate}) with typical CVs \citep{Xu+16}; 
(iv) Numerous CVs, including ones with massive white dwarfs \citep{Xu+19}, exist in the GC \citep{Zhu+18}, which ought to produce (recurrent-)novae observable in the X-ray band.
We elaborate on these points below.

A typical symbiotic nova hosts a red giant companion that drives a strong stellar wind, which can accumulate on the orbital plane under the gravitational influence of the binary system and form a density enhancement (DEOP; \citealp{Munari+25}).
Such a strong wind may also result in a relatively high mass-transfer rate, enhancing the likelihood of a recurrent nova.
During the outburst, material with a mass of approximately $10^{-7}$–-$10^{-4}~\mathrm{M_\odot}$ is ejected from the WD at velocities of $\sim10^3~\mathrm{km~s^{-1}}$ \citep{Starrfield+16}, colliding with the pre-existing DEOP and producing shocks with a broad temperature distribution.
The hottest components can reach $\gtrsim10$ keV, producing copious X-ray emission, a picture supported by X-ray observations \citep{Bode+06} and confirmed by hydrodynamic simulations \citep{Orlando+09}.

\begin{figure*}
    \centering
    \includegraphics[width=0.32\linewidth]{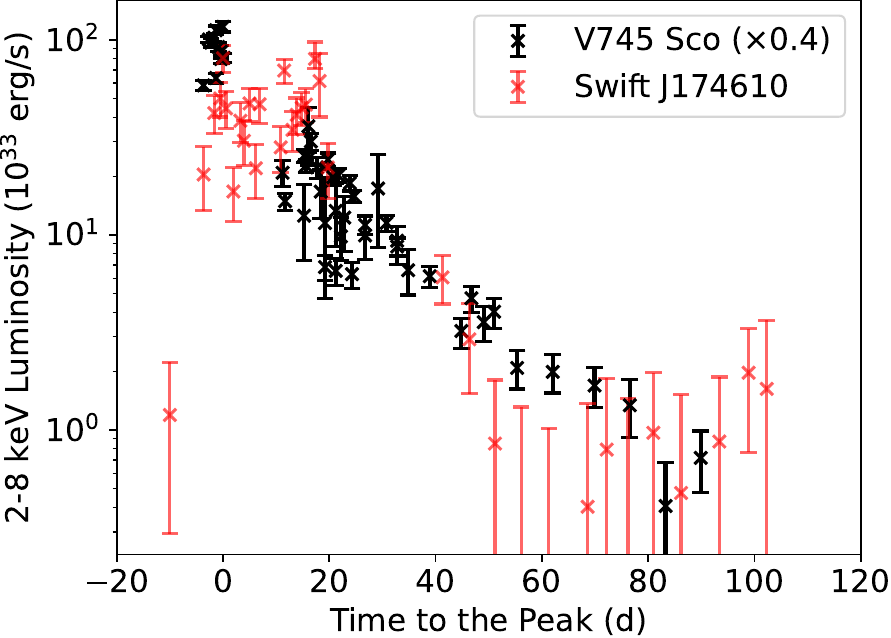}
    \includegraphics[width=0.32\linewidth]{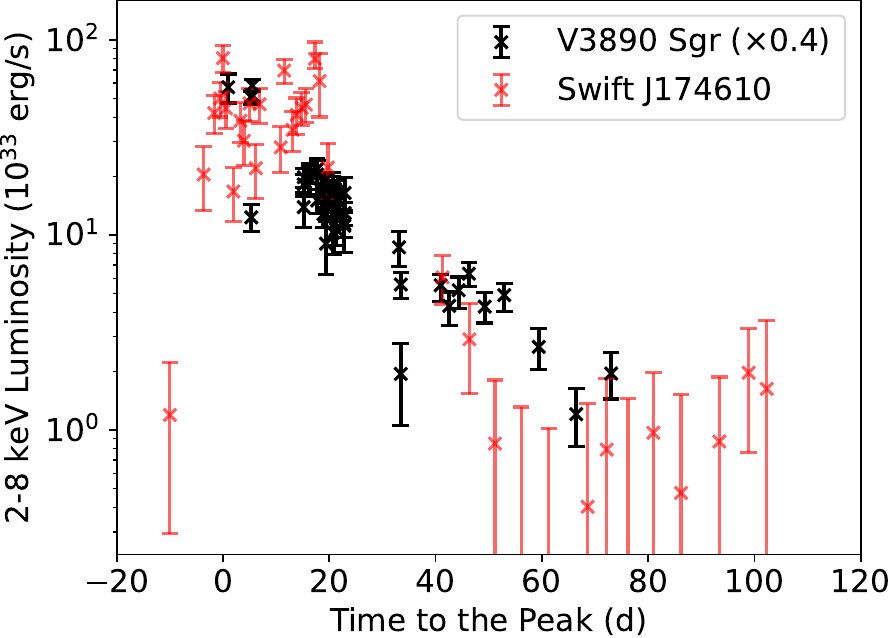}
    \includegraphics[width=0.32\linewidth]{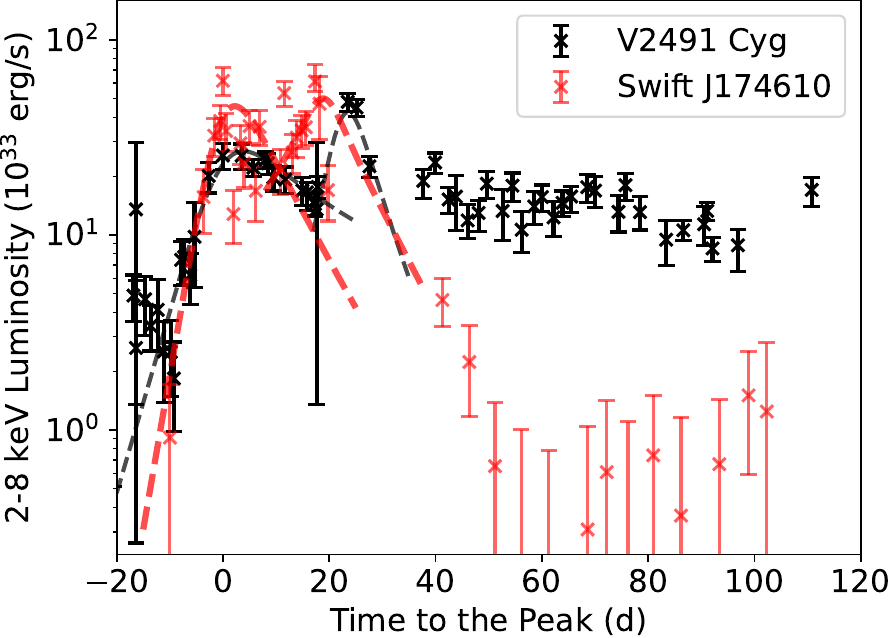}
    \caption{The 2–8 keV Swift/XRT light curves of V745 Sco ({\it left}), V3890 Sgr ({\it middle}), and V2491 Cyg ({\it right}) in black symbols, with the light curve of Swift J174610 (red symbols) overlaid for comparison.
    The luminosities are calculated based on count rates derived with the same procedure as Swift J174610 (available at \url{https://box.nju.edu.cn/d/effc773a4ddd48e48cd5/}).
    All light curves are aligned such that their first peak corresponds to time zero and a denoted scaling factor of the luminosity is applied, for a better visual comparison.
    In the right panel, dashed lines represent the possible presence of two peaks in the light curve, which are schematic only and do not correspond to fitted models. }
    % {\bf [could be helpful to use a luminosity scale and overlay Swift J174610. The exact MJD is not important here, better use a same x-aixs, e.g. use the first peak as time zero]}}
    \label{fig:nova_lc}
\end{figure*}

Currently, only four symbiotic nova systems have been observed to undergo multiple outbursts \citep{Darnley+21,RodriguezGil+23}, among which RS Oph is the best-studied example.
During its most recent outbursts in 2006 and 2021, RS Oph showed X-ray spectra with a peak luminosity of $\sim10^{36}~\rm{erg~s^{-1}}$, a plasma temperature of $\sim20$ keV but with emission lines out of collisional ionization equilibrium (CIE, \citealp{Sokoloski+06, Islam+24}), which are broadly similar to the characteristics of the XRISM spectrum of Swift J174610.
In particular, both sources show a discrepancy between the temperatures inferred from the Fe line ratios and from the bremsstrahlung continuum.
This behavior has been observed in several symbiotic novae  where the shocked plasma is known to deviate from CIE during the early phases of the outburst \citep{Ness+07,Orio+23,Islam+24}.
Such non-equilibrium ionization (NEI) effects provide a natural explanation for the anomalous Fe line ratios observed in Swift J174610.
In addition, the presence of a multi-temperature plasma, where different temperature components dominate the line and continuum emission, can further contribute to the discrepancy.
The 1–10 keV count-rate light curve of the recurrent nova RS Ophiuchi follows an exponential decay after the peak \citep{Page+22}. 
For a direct comparison with Swift J174610, we convert the count rate to flux in the 2–8 keV band assuming an absorbed APEC model with a plasma temperature of 10 keV and a foreground absorption of $N_H=10^{23}~{\rm cm^{-2}}$, and renormalize it to a distance of 8 kpc. 
The resulting light curve is shown in Fig.~\ref{fig:lc}, alongside the multi-mission monitoring of Swift J174610 ({\it Swift}, {\it XRISM}, {\it NuSTAR}, and {\it Chandra}), which reveals a broadly similar temporal evolution.
% The 1–10 keV light curve of RS Oph follows an exponential decay post-peak \citep{Page+22}, as shown in Fig.~\ref{fig:lc} for direct comparison with Swift J174610, whose multi-mission monitoring ({\it Swift}, {\it XRISM}, {\it NuSTAR}, and {\it Chandra}) reveals a broadly similar temporal evolution.

For comparison, we also examined the 2–8 keV light curves of two recent symbiotic recurrent novae, V745 Sco (2014; \citealp{Page+15}) and V3890 Sgr (2019; \citealp{Page+20}), as well as the classical nova V2491 Cyg (\citealp{Ragan+10}).
Using archival PC-mode {\it Swift} data, we derived their 2–8 keV light curves following the same procedure as described in Section~\ref{sec:swift}.
The luminosities were then calculated assuming distances of 7.8, 6.0, and 10.5 kpc \citep{Schaefer+10,Schaefer+09,Helton+08}, respectively.
%using archival {\it Swift} data processed and analyzed the same way as for Swift J174610 and RS Oph.
The two symbiotic novae, V745 Sco and V3890 Sgr, exhibit short-term hard X-ray variability within $\sim$20 days of the outburst, similar to that seen in Swift J174610 (left and middle panels of Fig.~\ref{fig:nova_lc}).
We suggest that such rapid hard X-ray variability may plausibly arise from inhomogeneities in the structure of the dynamical ejecta–originated plasma (DEOP), which can lead to localized and time-dependent shock heating.
RS Oph, and other Galactic RNe, also exhibit characteristic optical, UV and soft X-ray variability associated with the photosphere of the nova ejecta, which unfortunately is unobservable for any GC counterpart like Swift J174610.
A possible secondary rise is also present in the 2024 outburst of Swift J174610 (see the dotted lines in the right panel of Fig.~\ref{fig:nova_lc}). 
Interestingly, based on our analysis of archival {\it Swift} data, a similar secondary X-ray peak is found in V2491 Cyg, occurring about $\sim$20 days after the primary maximum. 
Notably, this secondary peak appears close to the peak of the supersoft (blackbody) emission (MJD$\sim54610$, \citealp{Page+10}), although its physical origin remains unclear.

It is worth noting that wide binaries such as symbiotic stars are vulnerable to dynamical disruption in very dense, high-velocity-dispersion environments like the NSC \citep{Heggie+75}.
Nevertheless, Swift J174610 lies at a projected distance of $\sim$15 pc from Sgr A*, where stellar densities and encounter rates are much lower than in the inner parsec and symbiotic binaries may survive on Gyr timescales \citep{Alexander+14}.

In conclusion, multiple independent observational clues consistently indicate that the peculiar GC transient Swift J174610 belongs to a nova outburst, possibly recurrent and symbiotic.
While this explanation can account for many, if not most, observed properties of Swift J174610, several issues remain. 
First, the Cr lines detected during the outburst spectra are unusual.
Even if they arise from cooler plasma components (e.g., cooling post-shock gas) than the dominant $\sim10$ keV component, reproducing the observed line strength would still require an extreme supersolar chromium abundance ($\gtrsim$5 times solar). 
Such an abundance cannot be produced by thermonuclear runaway on the surface of a WD, given the insufficient temperatures and densities involved \citep{Jose+98}.
Moreover, Cr lines have never been detected in the spectra of other GC X-ray sources, effectively ruling out a globally enhanced Cr abundance in this region.
One possibility is that chromium-rich dust grains, retained from previous outbursts in the DEOP, were destroyed during the current eruption, releasing Cr into the gas phase. 
In fact, evidence for supersolar Cr is not unprecedented in nova environments.
High-resolution optical spectroscopy has revealed transient heavy-element absorption (THEA) systems in which Fe-peak elements, including Cr, are enhanced above solar \citep{Williams+08}, indicating that elevated Cr may reflect pre-existing enriched circumbinary material rather than in-situ nucleosynthesis.

Second, the hard X-ray light curve around the peak of the 2024 outburst exhibits noticeable fluctuations that are absent in the outbursts of RS Ophiuchi (Fig.~\ref{fig:lc}).
This behavior is distinct from the commonly discussed rapid variability in the soft X-ray band.
Instead, it manifests as non-smooth, undulating flux variations in the hard (2–8 keV) X-ray light curve on short timescales.
Similar fluctuations are seen in V745 Sco and V3890 Sgr as well.
We find that the early shock expansion running into the inner portion of the DEOP can at least partly account for the hard X-ray fluctuations.
In this phase, the ejecta-driven shock propagates through a highly structured and inhomogeneous density field shaped by the prior wind–accretion evolution, including spiral-like features and local overdensities. 
As the shock encounters these density enhancements, the post-shock plasma is intermittently compressed and heated to temperatures of $\gtrsim10^8$ K, leading to episodic increases in the hard X-ray emissivity. 
Conversely, when the shock propagates into relatively tenuous regions, the emission temporarily weakens. 
This interaction between the shock front and the clumpy inner DEOP therefore naturally gives rise to short-timescale variability in the hard X-ray band during the early phase of the outburst.
This interaction between the shock front and the clumpy inner DEOP therefore naturally gives rise to short-timescale variability in the hard X-ray band during the early phase of the outburst. 
A more quantitative characterization of this scenario would require detailed, high-resolution hydrodynamical simulations, which is beyond the scope of this work.
The dust scattering in the foreground helps redistribute the intrinsic flux around the peak but cannot solely account for the observed degree of variability (see details in Appendix~\ref{sec:dust}).
On the other hand, the existence of a secondary peak may also add to the observed fluctuation.

Third, symbiotic stars typically exhibit a prominent Fe K$\alpha$ fluorescence line at 6.4 keV, produced when hard X-ray photons from the central source irradiate surrounding neutral or low-ionization material.
However, the line ratio of $I_{6.4}/I_{6.7} \sim 0.5$ observed during quiescence is significantly lower than that in several symbiotic systems ($\sim$2-3) studied by \citet{Xu+16}, distinguishing Swift J174610 from known symbiotic systems.

Fourth, the suggested existence of the 2005 outburst is not unambiguous, but this does not affect the need for a symbiotic nova outburst to explain the observed X-ray luminosity of the 2024 outburst, because classical novae typically have much lower peak X-ray luminosities due to the absence of a DEOP. 

%%%%%%%%%%%%%%%%%%%%%%%%%%%%%%%%%%%%%%%%%%%%%%%%%%
%%%%%%%%%%%%%%%%%%%%%%%%%%%%%%%%%%%%%%%%%%%%%%%%%%
%%%%%%%%%%%%%%%%%%%%%%%%%%%%%%%%%%%%%%%%%%%%%%%%%%

\section{Conclusion}
\label{sec:summary}
We have presented a comprehensive analysis of the transient source Swift J174610.4-290018, detected by {\it Swift} in February 2024 and subsequently observed with {\it Chandra}, {\it NuSTAR}, and {\it XRISM}, tracing a 2--8 keV luminosity declining from a peak luminosity of $\sim 10^{35}~\rm{erg~s^{-1}}$ to $\sim 10^{33}~\rm{erg~s^{-1}}$ over $\sim$120 days.
Examination of archival {\it Chandra} observations reveals a plausible past outburst in 2005, in addition to a quiescent state with a mean luminosity of $\sim 10^{32}~\rm{erg~s^{-1}}$ prevalent in the past 25 yrs.  
The X-ray spectra at all flux states exhibit prominent emission lines from helium- and hydrogen-like iron, as well as tentatively classified chromium lines.

The ADC scenario proposed by earlier work to explain the XRISM spectra is examined against the new information obtained in the present study, and is disfavored due to several points of inconsistency.

Instead, the temporal evolution, spectral characteristics, and similarities with the X-ray outbursts of RS Oph, and a few other Galactic novae, strongly support a nova origin for Swift J174610, although some issues remain to be understood, such as the presence of significant Cr lines and rapid flux variability seen in the first few days of the 2024 outburst.

If confirmed, Swift J174610 would be identified as the first detected (possibly recurrent and symbiotic) nova in the Galactic center, providing unambiguous evidence for the existence of wide binaries therein despite frequent stellar dynamical encounters, and important insights on the population of massive white dwarfs in the vicinity of Sgr A*. 
Multi-wavelength follow-up observations of Swift J174610 would be particularly interesting.

\section*{Acknowledgements}

This paper employs a list of Chandra data sets, obtained by the Chandra X-ray Observatory, which is contained in the Chandra Data Collection (CDC; \href{https://doi.org/10.25574/cdc.479}{DOI:10.25574/cdc.479}).
This work is supported by the National Natural Science Foundation of China (grant 12225302) and the Fundamental Research Funds for the Central Universities (grant KG202502).
The authors wish to thank Tong Bao, Eda Gjergo, Xing Lu, Fangzheng Shi and Qin Wu for helpful discussions.

\section*{Data Availability}
The data underlying this article will be shared on reasonable request to the corresponding author.
In addition, the 2-8 keV count rate lightcurves of V745 Sco, V3890 Sgr and V2491 Cyg are publicly available at \href{https://box.nju.edu.cn/d/effc773a4ddd48e48cd5/}{NJU box}.



\bibliographystyle{mnras}
\bibliography{main} % if your bibtex file is called example.bib

@ARTICLE{Genzel+10,
       author = {{Genzel}, Reinhard and {Eisenhauer}, Frank and {Gillessen}, Stefan},
        title = "{The Galactic Center massive black hole and nuclear star cluster}",
      journal = {Reviews of Modern Physics},
         year = 2010,
        month = oct,
       volume = {82},
       number = {4},
        pages = {3121-3195},
          doi = {10.1103/RevModPhys.82.3121},
archivePrefix = {arXiv},
       eprint = {1006.0064},
 primaryClass = {astro-ph.GA},
       adsurl = {https://ui.adsabs.harvard.edu/abs/2010RvMP...82.3121G}
}

@ARTICLE{Do+19,
       author = {{Do}, Tuan and {Witzel}, Gunther and {Gautam}, Abhimat K. and {Chen}, Zhuo and {Ghez}, Andrea M. and {Morris}, Mark R. and {Becklin}, Eric E. and {Ciurlo}, Anna and {Hosek}, Matthew, Jr. and {Martinez}, Gregory D. and {Matthews}, Keith and {Sakai}, Shoko and {Sch{\"o}del}, Rainer},
        title = "{Unprecedented Near-infrared Brightness and Variability of Sgr A*}",
      journal = {\apjl},
         year = 2019,
        month = sep,
       volume = {882},
       number = {2},
          eid = {L27},
        pages = {L27},
          doi = {10.3847/2041-8213/ab38c3},
archivePrefix = {arXiv},
       eprint = {1908.01777},
 primaryClass = {astro-ph.GA},
       adsurl = {https://ui.adsabs.harvard.edu/abs/2019ApJ...882L..27D}
}

@ARTICLE{Gravity_collaboration+20,
       author = {{Gravity Collaboration} and {Abuter}, R. and {Amorim}, A. and {Baub{\"o}ck}, M. and {Berger}, J.~P. and {Bonnet}, H. and {Brandner}, W. and {Cardoso}, V. and {Cl{\'e}net}, Y. and {de Zeeuw}, P.~T. and {Dexter}, J. and {Eckart}, A. and {Eisenhauer}, F. and {F{\"o}rster Schreiber}, N.~M. and {Garcia}, P. and {Gao}, F. and {Gendron}, E. and {Genzel}, R. and {Gillessen}, S. and {Habibi}, M. and {Haubois}, X. and {Henning}, T. and {Hippler}, S. and {Horrobin}, M. and {Jim{\'e}nez-Rosales}, A. and {Jochum}, L. and {Jocou}, L. and {Kaufer}, A. and {Kervella}, P. and {Lacour}, S. and {Lapeyr{\`e}re}, V. and {Le Bouquin}, J. -B. and {L{\'e}na}, P. and {Nowak}, M. and {Ott}, T. and {Paumard}, T. and {Perraut}, K. and {Perrin}, G. and {Pfuhl}, O. and {Rodr{\'\i}guez-Coira}, G. and {Shangguan}, J. and {Scheithauer}, S. and {Stadler}, J. and {Straub}, O. and {Straubmeier}, C. and {Sturm}, E. and {Tacconi}, L.~J. and {Vincent}, F. and {von Fellenberg}, S. and {Waisberg}, I. and {Widmann}, F. and {Wieprecht}, E. and {Wiezorrek}, E. and {Woillez}, J. and {Yazici}, S. and {Zins}, G.},
        title = "{Detection of the Schwarzschild precession in the orbit of the star S2 near the Galactic centre massive black hole}",
      journal = {\aap},
         year = 2020,
        month = apr,
       volume = {636},
          eid = {L5},
        pages = {L5},
          doi = {10.1051/0004-6361/202037813},
archivePrefix = {arXiv},
       eprint = {2004.07187},
 primaryClass = {astro-ph.GA},
       adsurl = {https://ui.adsabs.harvard.edu/abs/2020A&A...636L...5G}
}

@ARTICLE{Feldmeier-Krause+17,
       author = {{Feldmeier-Krause}, A. and {Zhu}, L. and {Neumayer}, N. and {van de Ven}, G. and {de Zeeuw}, P.~T. and {Sch{\"o}del}, R.},
        title = "{Triaxial orbit-based modelling of the Milky Way Nuclear Star Cluster}",
      journal = {\mnras},
         year = 2017,
        month = apr,
       volume = {466},
       number = {4},
        pages = {4040-4052},
          doi = {10.1093/mnras/stw3377},
archivePrefix = {arXiv},
       eprint = {1701.01583},
 primaryClass = {astro-ph.GA},
       adsurl = {https://ui.adsabs.harvard.edu/abs/2017MNRAS.466.4040F}
}

@ARTICLE{Launhardt+02,
       author = {{Launhardt}, R. and {Zylka}, R. and {Mezger}, P.~G.},
        title = "{The nuclear bulge of the Galaxy. III. Large-scale physical characteristics of stars and interstellar matter}",
      journal = {\aap},
         year = 2002,
        month = mar,
       volume = {384},
        pages = {112-139},
          doi = {10.1051/0004-6361:20020017},
archivePrefix = {arXiv},
       eprint = {astro-ph/0201294},
 primaryClass = {astro-ph},
       adsurl = {https://ui.adsabs.harvard.edu/abs/2002A&A...384..112L}
}

@ARTICLE{Sormani+22,
       author = {{Sormani}, Mattia C. and {Sanders}, Jason L. and {Fritz}, Tobias K. and {Smith}, Leigh C. and {Gerhard}, Ortwin and {Sch{\"o}del}, Rainer and {Magorrian}, John and {Neumayer}, Nadine and {Nogueras-Lara}, Francisco and {Feldmeier-Krause}, Anja and {Mastrobuono-Battisti}, Alessandra and {Schultheis}, Mathias and {Shahzamanian}, Banafsheh and {Vasiliev}, Eugene and {Klessen}, Ralf S. and {Lucas}, Philip and {Minniti}, Dante},
        title = "{Self-consistent modelling of the Milky Way's nuclear stellar disc}",
      journal = {\mnras},
         year = 2022,
        month = may,
       volume = {512},
       number = {2},
        pages = {1857-1884},
          doi = {10.1093/mnras/stac639},
archivePrefix = {arXiv},
       eprint = {2111.12713},
 primaryClass = {astro-ph.GA},
       adsurl = {https://ui.adsabs.harvard.edu/abs/2022MNRAS.512.1857S}
}

@ARTICLE{Weisskopf+02,
       author = {{Weisskopf}, M.~C. and {Brinkman}, B. and {Canizares}, C. and {Garmire}, G. and {Murray}, S. and {Van Speybroeck}, L.~P.},
        title = "{An Overview of the Performance and Scientific Results from the Chandra X-Ray Observatory}",
      journal = {\pasp},
         year = 2002,
        month = jan,
       volume = {114},
       number = {791},
        pages = {1-24},
          doi = {10.1086/338108},
archivePrefix = {arXiv},
       eprint = {astro-ph/0110308},
 primaryClass = {astro-ph},
       adsurl = {https://ui.adsabs.harvard.edu/abs/2002PASP..114....1W}
}

@ARTICLE{Wang+02,
       author = {{Wang}, Q.~D. and {Gotthelf}, E.~V. and {Lang}, C.~C.},
        title = "{A faint discrete source origin for the highly ionized iron emission from the Galactic Centre region}",
      journal = {\nat},
         year = 2002,
        month = jan,
       volume = {415},
       number = {6868},
        pages = {148-150},
          doi = {10.1038/415148a},
       adsurl = {https://ui.adsabs.harvard.edu/abs/2002Natur.415..148W}
}

@ARTICLE{Muno+03,
       author = {{Muno}, M.~P. and {Baganoff}, F.~K. and {Bautz}, M.~W. and {Brandt}, W.~N. and {Broos}, P.~S. and {Feigelson}, E.~D. and {Garmire}, G.~P. and {Morris}, M.~R. and {Ricker}, G.~R. and {Townsley}, L.~K.},
        title = "{A Deep Chandra Catalog of X-Ray Point Sources toward the Galactic Center}",
      journal = {\apj},
         year = 2003,
        month = may,
       volume = {589},
       number = {1},
        pages = {225-241},
          doi = {10.1086/374639},
archivePrefix = {arXiv},
       eprint = {astro-ph/0301371},
 primaryClass = {astro-ph},
       adsurl = {https://ui.adsabs.harvard.edu/abs/2003ApJ...589..225M}
}

@ARTICLE{Muno+06,
       author = {{Muno}, M.~P. and {Bauer}, F.~E. and {Bandyopadhyay}, R.~M. and {Wang}, Q.~D.},
        title = "{A Chandra Catalog of X-Ray Sources in the Central 150 pc of the Galaxy}",
      journal = {\apjs},
         year = 2006,
        month = jul,
       volume = {165},
       number = {1},
        pages = {173-187},
          doi = {10.1086/504798},
archivePrefix = {arXiv},
       eprint = {astro-ph/0601627},
 primaryClass = {astro-ph},
       adsurl = {https://ui.adsabs.harvard.edu/abs/2006ApJS..165..173M}
}

@ARTICLE{Muno+09,
       author = {{Muno}, M.~P. and {Bauer}, F.~E. and {Baganoff}, F.~K. and {Bandyopadhyay}, R.~M. and {Bower}, G.~C. and {Brandt}, W.~N. and {Broos}, P.~S. and {Cotera}, A. and {Eikenberry}, S.~S. and {Garmire}, G.~P. and {Hyman}, S.~D. and {Kassim}, N.~E. and {Lang}, C.~C. and {Lazio}, T.~J.~W. and {Law}, C. and {Mauerhan}, J.~C. and {Morris}, M.~R. and {Nagata}, T. and {Nishiyama}, S. and {Park}, S. and {Ram{\`\i}rez}, S.~V. and {Stolovy}, S.~R. and {Wijnands}, R. and {Wang}, Q.~D. and {Wang}, Z. and {Yusef-Zadeh}, F.},
        title = "{A Catalog of X-Ray Point Sources from Two Megaseconds of Chandra Observations of the Galactic Center}",
      journal = {\apjs},
         year = 2009,
        month = mar,
       volume = {181},
       number = {1},
        pages = {110-128},
          doi = {10.1088/0067-0049/181/1/110},
archivePrefix = {arXiv},
       eprint = {0809.1105},
 primaryClass = {astro-ph},
       adsurl = {https://ui.adsabs.harvard.edu/abs/2009ApJS..181..110M}
}

@ARTICLE{Zhu+18,
       author = {{Zhu}, Zhenlin and {Li}, Zhiyuan and {Morris}, Mark R.},
        title = "{An Ultradeep Chandra Catalog of X-Ray Point Sources in the Galactic Center Star Cluster}",
      journal = {\apjs},
         year = 2018,
        month = apr,
       volume = {235},
       number = {2},
          eid = {26},
        pages = {26},
          doi = {10.3847/1538-4365/aab14f},
archivePrefix = {arXiv},
       eprint = {1802.05073},
 primaryClass = {astro-ph.HE},
       adsurl = {https://ui.adsabs.harvard.edu/abs/2018ApJS..235...26Z}
}

@ARTICLE{Degenaar+12,
       author = {{Degenaar}, N. and {Wijnands}, R. and {Cackett}, E.~M. and {Homan}, J. and {in't Zand}, J.~J.~M. and {Kuulkers}, E. and {Maccarone}, T.~J. and {van der Klis}, M.},
        title = "{A four-year XMM-Newton/Chandra monitoring campaign of the Galactic centre: analysing the X-ray transients}",
      journal = {\aap},
         year = 2012,
        month = sep,
       volume = {545},
          eid = {A49},
        pages = {A49},
          doi = {10.1051/0004-6361/201219470},
archivePrefix = {arXiv},
       eprint = {1204.6043},
 primaryClass = {astro-ph.HE},
       adsurl = {https://ui.adsabs.harvard.edu/abs/2012A&A...545A..49D}
}

@ARTICLE{Ponti+16,
       author = {{Ponti}, G. and {Jin}, C. and {De Marco}, B. and {Rea}, N. and {Rau}, A. and {Haberl}, F. and {Coti Zelati}, F. and {Bozzo}, E. and {Ferrigno}, C. and {Bower}, G.~C. and {Demorest}, P.},
        title = "{Swift J174540.7-290015: a new accreting binary in the Galactic Centre}",
      journal = {\mnras},
         year = 2016,
        month = sep,
       volume = {461},
       number = {3},
        pages = {2688-2701},
          doi = {10.1093/mnras/stw1382},
archivePrefix = {arXiv},
       eprint = {1606.01138},
 primaryClass = {astro-ph.HE},
       adsurl = {https://ui.adsabs.harvard.edu/abs/2016MNRAS.461.2688P}
}

@ARTICLE{Degenaar+15,
       author = {{Degenaar}, N. and {Wijnands}, R. and {Miller}, J.~M. and {Reynolds}, M.~T. and {Kennea}, J. and {Gehrels}, N.},
        title = "{The Swift X-ray monitoring campaign of the center of the Milky Way}",
      journal = {Journal of High Energy Astrophysics},
         year = 2015,
        month = sep,
       volume = {7},
        pages = {137-147},
          doi = {10.1016/j.jheap.2015.03.005},
archivePrefix = {arXiv},
       eprint = {1503.07524},
 primaryClass = {astro-ph.HE},
       adsurl = {https://ui.adsabs.harvard.edu/abs/2015JHEAp...7..137D}
}

@INCOLLECTION{Bahramian+23,
       author = {{Bahramian}, Arash and {Degenaar}, Nathalie},
        title = "{Low-Mass X-ray Binaries}",
    booktitle = {Handbook of X-ray and Gamma-ray Astrophysics},
         year = 2023,
          eid = {120},
        pages = {120},
          doi = {10.1007/978-981-16-4544-0_94-1},
       adsurl = {https://ui.adsabs.harvard.edu/abs/2023hxga.book..120B}
}

@ARTICLE{Wijnands+06,
       author = {{Wijnands}, R. and {in't Zand}, J.~J.~M. and {Rupen}, M. and {Maccarone}, T. and {Homan}, J. and {Cornelisse}, R. and {Fender}, R. and {Grindlay}, J. and {van der Klis}, M. and {Kuulkers}, E. and {Markwardt}, C.~B. and {Miller-Jones}, J.~C.~A. and {Wang}, Q.~D.},
        title = "{The XMM-Newton/Chandra monitoring campaign of the Galactic center region. Description of the program and preliminary results}",
      journal = {\aap},
         year = 2006,
        month = apr,
       volume = {449},
       number = {3},
        pages = {1117-1127},
          doi = {10.1051/0004-6361:20054129},
archivePrefix = {arXiv},
       eprint = {astro-ph/0508648},
 primaryClass = {astro-ph},
       adsurl = {https://ui.adsabs.harvard.edu/abs/2006A&A...449.1117W}
}

@ARTICLE{Reynolds+24,
       author = {{Reynolds}, Mark and {Degenaar}, Natalie and {Wijnands}, Rudy and {Miller}, Jon and {Kennea}, Jamie},
        title = "{Swift GC monitoring program detection of low luminosity outburst a new source: Swift J174610-290018}",
      journal = {The Astronomer's Telegram},
         year = 2024,
        month = feb,
       volume = {16481},
        pages = {1},
       adsurl = {https://ui.adsabs.harvard.edu/abs/2024ATel16481....1R}
}

@ARTICLE{Yoshimoto+25,
       author = {{Yoshimoto}, Anje and {Yamauchi}, Shigeo and {Nobukawa}, Masayoshi and {Uchiyama}, Hideki and {Nobukawa}, Kumiko K. and {Aoki}, Yuma and {Ishida}, Manabu and {Kanemaru}, Yoshiaki and {Shidatsu}, Megumi and {Hayashi}, Takayuki and {Maeda}, Yoshitomo and {Matsumoto}, Hironori and {Tsuboi}, Yohko and {Suzuki}, Hiromasa and {Nakajima}, Hiroshi and {Wang}, Q. Daniel and {Eguchi}, Satoshi and {Yoneyama}, Tomokage and {Dotani}, Tadayasu and {Behar}, Ehud and {Terada}, Yukikatsu and {Suzuki}, Nari and {Yoshimoto}, Marina},
        title = "{The unusual spectrum of the X-ray transient source XRISM J174610.8-290021 near the Galactic Center}",
      journal = {\pasj},
         year = 2025,
        month = jun,
          doi = {10.1093/pasj/psaf063},
archivePrefix = {arXiv},
       eprint = {2506.20088},
 primaryClass = {astro-ph.HE},
       adsurl = {https://ui.adsabs.harvard.edu/abs/2025PASJ..tmp...74Y}
}

@ARTICLE{Iaria+13,
       author = {{Iaria}, R. and {Di Salvo}, T. and {D'A{\`\i}}, A. and {Burderi}, L. and {Mineo}, T. and {Riggio}, A. and {Papitto}, A. and {Robba}, N.~R.},
        title = "{X-ray spectroscopy of the ADC source X1822-371 with Chandra and XMM-Newton}",
      journal = {\aap},
         year = 2013,
        month = jan,
       volume = {549},
          eid = {A33},
        pages = {A33},
          doi = {10.1051/0004-6361/201015305},
archivePrefix = {arXiv},
       eprint = {1210.0874},
 primaryClass = {astro-ph.HE},
       adsurl = {https://ui.adsabs.harvard.edu/abs/2013A&A...549A..33I}
}

@INPROCEEDINGS{Moretti+05,
       author = {{Moretti}, Alberto and {Campana}, Sergio and {Mineo}, T. and {Romano}, Patrizia and {Abbey}, A.~F. and {Angelini}, L. and {Beardmore}, A. and {Burkert}, W. and {Burrows}, D.~N. and {Capalbi}, M. and {Chincarini}, G. and {Citterio}, O. and {Cusumano}, G. and {Freyberg}, M.~J. and {Giommi}, P. and {Goad}, M.~R. and {Godet}, O. and {Hartner}, G.~D. and {Hill}, J.~E. and {Kennea}, J. and {La Parola}, V. and {Mangano}, V. and {Morris}, D. and {Nousek}, J.~A. and {Osborne}, J. and {Page}, K. and {Pagani}, C. and {Perri}, M. and {Tagliaferri}, G. and {Tamburelli}, F. and {Wells}, A.},
        title = "{In-flight calibration of the Swift XRT Point Spread Function}",
    booktitle = {UV, X-Ray, and Gamma-Ray Space Instrumentation for Astronomy XIV},
         year = 2005,
       editor = {{Siegmund}, Oswald H.~W.},
       series = {Society of Photo-Optical Instrumentation Engineers (SPIE) Conference Series},
       volume = {5898},
        month = aug,
        pages = {360-368},
          doi = {10.1117/12.617164},
       adsurl = {https://ui.adsabs.harvard.edu/abs/2005SPIE.5898..360M}
}

@ARTICLE{White+82,
       author = {{White}, N.~E. and {Holt}, S.~S.},
        title = "{Accretion disk coronae.}",
      journal = {\apj},
         year = 1982,
        month = jun,
       volume = {257},
        pages = {318-337},
          doi = {10.1086/159991},
       adsurl = {https://ui.adsabs.harvard.edu/abs/1982ApJ...257..318W}
}

@ARTICLE{Yoneyama+23,
       author = {{Yoneyama}, Tomokage and {Dotani}, Tadayasu},
        title = "{X-ray spectroscopy of the accretion disk corona source 2S 0921-630 with Suzaku archival data}",
      journal = {\pasj},
         year = 2023,
        month = feb,
       volume = {75},
       number = {1},
        pages = {30-36},
          doi = {10.1093/pasj/psac086},
archivePrefix = {arXiv},
       eprint = {2210.10792},
 primaryClass = {astro-ph.HE},
       adsurl = {https://ui.adsabs.harvard.edu/abs/2023PASJ...75...30Y}
}

@ARTICLE{Nowak+02,
       author = {{Nowak}, Michael A. and {Heinz}, Sebastian and {Begelman}, M.~C.},
        title = "{Hiding in Plain Sight: Chandra Observations of the Quiescent Neutron Star 4U 2129+47 in Eclipse}",
      journal = {\apj},
         year = 2002,
        month = jul,
       volume = {573},
       number = {2},
        pages = {778-788},
          doi = {10.1086/340757},
archivePrefix = {arXiv},
       eprint = {astro-ph/0204503},
 primaryClass = {astro-ph},
       adsurl = {https://ui.adsabs.harvard.edu/abs/2002ApJ...573..778N}
}

@ARTICLE{Done+07,
       author = {{Done}, Chris and {Gierli{\'n}ski}, Marek and {Kubota}, Aya},
        title = "{Modelling the behaviour of accretion flows in X-ray binaries. Everything you always wanted to know about accretion but were afraid to ask}",
      journal = {\aapr},
         year = 2007,
        month = dec,
       volume = {15},
       number = {1},
        pages = {1-66},
          doi = {10.1007/s00159-007-0006-1},
archivePrefix = {arXiv},
       eprint = {0708.0148},
 primaryClass = {astro-ph},
       adsurl = {https://ui.adsabs.harvard.edu/abs/2007A&ARv..15....1D}
}

@ARTICLE{Church+04,
       author = {{Church}, M.~J. and {Ba{\l}uci{\'n}ska-Church}, M.},
        title = "{Measurements of accretion disc corona size in LMXB: consequences for Comptonization and LMXB models}",
      journal = {\mnras},
         year = 2004,
        month = mar,
       volume = {348},
       number = {3},
        pages = {955-963},
          doi = {10.1111/j.1365-2966.2004.07162.x},
archivePrefix = {arXiv},
       eprint = {astro-ph/0309212},
 primaryClass = {astro-ph},
       adsurl = {https://ui.adsabs.harvard.edu/abs/2004MNRAS.348..955C}
}

@ARTICLE{Jin+17,
       author = {{Jin}, Chichuan and {Ponti}, Gabriele and {Haberl}, Frank and {Smith}, Randall},
        title = "{Probing the interstellar dust towards the Galactic Centre: dust-scattering halo around AX J1745.6-2901}",
      journal = {\mnras},
         year = 2017,
        month = jul,
       volume = {468},
       number = {3},
        pages = {2532-2551},
          doi = {10.1093/mnras/stx653},
archivePrefix = {arXiv},
       eprint = {1703.05179},
 primaryClass = {astro-ph.HE},
       adsurl = {https://ui.adsabs.harvard.edu/abs/2017MNRAS.468.2532J}
}

@ARTICLE{Hua+25,
       author = {{Hua}, Ziqian and {Li}, Zhiyuan},
        title = "{Chandra X-ray measurement of heavy element abundances of Wolf{\textendash}Rayet stars in the Galactic Centre}",
      journal = {\mnras},
         year = 2025,
        month = jul,
       volume = {540},
       number = {4},
        pages = {3850-3862},
          doi = {10.1093/mnras/staf957},
archivePrefix = {arXiv},
       eprint = {2506.08587},
 primaryClass = {astro-ph.GA},
       adsurl = {https://ui.adsabs.harvard.edu/abs/2025MNRAS.540.3850H}
}

@ARTICLE{Mathis+91,
       author = {{Mathis}, John S. and {Lee}, C. -W.},
        title = "{X-Ray Halos as Diagnostics of Interstellar Grains}",
      journal = {\apj},
         year = 1991,
        month = aug,
       volume = {376},
        pages = {490},
          doi = {10.1086/170297},
       adsurl = {https://ui.adsabs.harvard.edu/abs/1991ApJ...376..490M}
}

@ARTICLE{Starrfield+16,
       author = {{Starrfield}, S. and {Iliadis}, C. and {Hix}, W.~R.},
        title = "{The Thermonuclear Runaway and the Classical Nova Outburst}",
      journal = {\pasp},
         year = 2016,
        month = may,
       volume = {128},
       number = {963},
        pages = {051001},
          doi = {10.1088/1538-3873/128/963/051001},
archivePrefix = {arXiv},
       eprint = {1605.04294},
 primaryClass = {astro-ph.SR},
       adsurl = {https://ui.adsabs.harvard.edu/abs/2016PASP..128e1001S}
}

@ARTICLE{Munari+25,
       author = {{Munari}, U.},
        title = "{Symbiotic novae}",
      journal = {Contributions of the Astronomical Observatory Skalnate Pleso},
         year = 2025,
        month = apr,
       volume = {55},
       number = {3},
        pages = {47-66},
          doi = {10.31577/caosp.2025.55.3.47},
archivePrefix = {arXiv},
       eprint = {2412.20499},
 primaryClass = {astro-ph.SR},
       adsurl = {https://ui.adsabs.harvard.edu/abs/2025CoSka..55c..47M}
}

@INPROCEEDINGS{Darnley+21,
       author = {{Darnley}, M.~J.},
        title = "{Accrete, Accrete, Accrete{\textellipsis} Bang! (and repeat): The remarkable Recurrent Novae}",
    booktitle = {The Golden Age of Cataclysmic Variables and Related Objects V},
         year = 2021,
       volume = {2-7},
        month = feb,
          eid = {44},
        pages = {44},
          doi = {10.22323/1.368.0044},
archivePrefix = {arXiv},
       eprint = {1912.13209},
 primaryClass = {astro-ph.SR},
       adsurl = {https://ui.adsabs.harvard.edu/abs/2021gacv.workE..44D}
}

@ARTICLE{Islam+24,
       author = {{Islam}, Nazma and {Mukai}, Koji and {Sokoloski}, J.~L.},
        title = "{X-Rays from RS Ophiuchi's 2021 Eruption: Shocks In and Out of Ionization Equilibrium}",
      journal = {\apj},
         year = 2024,
        month = jan,
       volume = {960},
       number = {2},
          eid = {125},
        pages = {125},
          doi = {10.3847/1538-4357/ad1041},
archivePrefix = {arXiv},
       eprint = {2311.17156},
 primaryClass = {astro-ph.HE},
       adsurl = {https://ui.adsabs.harvard.edu/abs/2024ApJ...960..125I}
}

@ARTICLE{Sokoloski+06,
       author = {{Sokoloski}, J.~L. and {Luna}, G.~J.~M. and {Mukai}, K. and {Kenyon}, Scott J.},
        title = "{An X-ray-emitting blast wave from the recurrent nova RS Ophiuchi}",
      journal = {\nat},
         year = 2006,
        month = jul,
       volume = {442},
       number = {7100},
        pages = {276-278},
          doi = {10.1038/nature04893},
archivePrefix = {arXiv},
       eprint = {astro-ph/0605326},
 primaryClass = {astro-ph},
       adsurl = {https://ui.adsabs.harvard.edu/abs/2006Natur.442..276S}
}

@ARTICLE{Page+22,
       author = {{Page}, K.~L. and {Beardmore}, A.~P. and {Osborne}, J.~P. and {Munari}, U. and {Ness}, J. -U. and {Evans}, P.~A. and {Bode}, M.~F. and {Darnley}, M.~J. and {Drake}, J.~J. and {Kuin}, N.~P.~M. and {O'Brien}, T.~J. and {Orio}, M. and {Shore}, S.~N. and {Starrfield}, S. and {Woodward}, C.~E.},
        title = "{The 2021 outburst of the recurrent nova RS Ophiuchi observed in X-rays by the Neil Gehrels Swift Observatory: a comparative study}",
      journal = {\mnras},
         year = 2022,
        month = aug,
       volume = {514},
       number = {2},
        pages = {1557-1574},
          doi = {10.1093/mnras/stac1295},
archivePrefix = {arXiv},
       eprint = {2205.03232},
 primaryClass = {astro-ph.HE},
       adsurl = {https://ui.adsabs.harvard.edu/abs/2022MNRAS.514.1557P}
}

@ARTICLE{Heggie+75,
       author = {{Heggie}, D.~C.},
        title = "{Binary evolution in stellar dynamics.}",
      journal = {\mnras},
         year = 1975,
        month = dec,
       volume = {173},
        pages = {729-787},
          doi = {10.1093/mnras/173.3.729},
       adsurl = {https://ui.adsabs.harvard.edu/abs/1975MNRAS.173..729H}
}

@ARTICLE{Alexander+14,
       author = {{Alexander}, Tal and {Pfuhl}, Oliver},
        title = "{Constraining the Dark Cusp in the Galactic Center by Long-period Binaries}",
      journal = {\apj},
         year = 2014,
        month = jan,
       volume = {780},
       number = {2},
          eid = {148},
        pages = {148},
          doi = {10.1088/0004-637X/780/2/148},
archivePrefix = {arXiv},
       eprint = {1308.6638},
 primaryClass = {astro-ph.GA},
       adsurl = {https://ui.adsabs.harvard.edu/abs/2014ApJ...780..148A}
}

@ARTICLE{Xu+16,
       author = {{Xu}, Xiao-jie and {Wang}, Q. Daniel and {Li}, Xiang-Dong},
        title = "{Fe Line Diagnostics of Cataclysmic Variables and Galactic Ridge X-Ray Emission}",
      journal = {\apj},
         year = 2016,
        month = feb,
       volume = {818},
       number = {2},
          eid = {136},
        pages = {136},
          doi = {10.3847/0004-637X/818/2/136},
archivePrefix = {arXiv},
       eprint = {1602.05262},
 primaryClass = {astro-ph.HE},
       adsurl = {https://ui.adsabs.harvard.edu/abs/2016ApJ...818..136X}
}

@ARTICLE{Jose+98,
       author = {{Jos{\'e}}, Jordi and {Hernanz}, Margarita},
        title = "{Nucleosynthesis in Classical Novae: CO versus ONe White Dwarfs}",
      journal = {\apj},
         year = 1998,
        month = feb,
       volume = {494},
       number = {2},
        pages = {680-690},
          doi = {10.1086/305244},
archivePrefix = {arXiv},
       eprint = {astro-ph/9709153},
 primaryClass = {astro-ph},
       adsurl = {https://ui.adsabs.harvard.edu/abs/1998ApJ...494..680J}
}

@INPROCEEDINGS{Arnaud+96,
       author = {{Arnaud}, K.~A.},
        title = "{XSPEC: The First Ten Years}",
    booktitle = {Astronomical Data Analysis Software and Systems V},
         year = 1996,
       editor = {{Jacoby}, George H. and {Barnes}, Jeannette},
       series = {Astronomical Society of the Pacific Conference Series},
       volume = {101},
        month = jan,
        pages = {17},
       adsurl = {https://ui.adsabs.harvard.edu/abs/1996ASPC..101...17A}
}

@ARTICLE{Boese+01,
       author = {{Boese}, F.~G. and {Doebereiner}, S.},
        title = "{Maximum likelihood estimation of single X-ray point-source parameters in ROSAT data}",
      journal = {\aap},
         year = 2001,
        month = may,
       volume = {370},
        pages = {649-671},
          doi = {10.1051/0004-6361:20010092},
       adsurl = {https://ui.adsabs.harvard.edu/abs/2001A&A...370..649B}
}

@ARTICLE{Bode+06,
       author = {{Bode}, M.~F. and {O'Brien}, T.~J. and {Osborne}, J.~P. and {Page}, K.~L. and {Senziani}, F. and {Skinner}, G.~K. and {Starrfield}, S. and {Ness}, J. -U. and {Drake}, J.~J. and {Schwarz}, G. and {Beardmore}, A.~P. and {Darnley}, M.~J. and {Eyres}, S.~P.~S. and {Evans}, A. and {Gehrels}, N. and {Goad}, M.~R. and {Jean}, P. and {Krautter}, J. and {Novara}, G.},
        title = "{Swift Observations of the 2006 Outburst of the Recurrent Nova RS Ophiuchi. I. Early X-Ray Emission from the Shocked Ejecta and Red Giant Wind}",
      journal = {\apj},
         year = 2006,
        month = nov,
       volume = {652},
       number = {1},
        pages = {629-635},
          doi = {10.1086/507980},
archivePrefix = {arXiv},
       eprint = {astro-ph/0604618},
 primaryClass = {astro-ph},
       adsurl = {https://ui.adsabs.harvard.edu/abs/2006ApJ...652..629B}
}

@ARTICLE{Orlando+09,
       author = {{Orlando}, S. and {Drake}, J.~J. and {Laming}, J.~M.},
        title = "{Three-dimensional modeling of the asymmetric blast wave from the 2006 outburst of RS Ophiuchi: Early X-ray emission}",
      journal = {\aap},
         year = 2009,
        month = jan,
       volume = {493},
       number = {3},
        pages = {1049-1059},
          doi = {10.1051/0004-6361:200810109},
archivePrefix = {arXiv},
       eprint = {0811.3941},
 primaryClass = {astro-ph},
       adsurl = {https://ui.adsabs.harvard.edu/abs/2009A&A...493.1049O}
}

@ARTICLE{Xu+19,
       author = {{Xu}, Xiao-jie and {Li}, Zhiyuan and {Zhu}, Zhenlin and {Cheng}, Zhongqun and {Li}, Xiang-dong and {Yu}, Zhuo-li},
        title = "{Massive White Dwarfs in the Galactic Center: A Chandra X-Ray Spectroscopy of Cataclysmic Variables}",
      journal = {\apj},
         year = 2019,
        month = sep,
       volume = {882},
       number = {2},
          eid = {164},
        pages = {164},
          doi = {10.3847/1538-4357/ab32df},
archivePrefix = {arXiv},
       eprint = {1907.09086},
 primaryClass = {astro-ph.HE},
       adsurl = {https://ui.adsabs.harvard.edu/abs/2019ApJ...882..164X}
}

@INPROCEEDINGS{Ragan+10,
       author = {{Ragan}, E. and {Mikol{\l}ajewski}, M. and {Tomov}, T. and {Swierczy{\'n}ski}, E. and {Bro{\.z}ek}, T. and {Ga{\l}an}, C. and {R{\'o}{\.z}a{\'n}ski}, P. and {Wiecek}, M. and {Wychudzki}, P.},
        title = "{V2491 Cyg - A Possible Recurrent Nova?}",
    booktitle = {Binaries - Key to Comprehension of the Universe},
         year = 2010,
       editor = {{Pr{\v{s}}a}, A. and {Zejda}, M.},
       series = {Astronomical Society of the Pacific Conference Series},
       volume = {435},
        month = dec,
        pages = {335},
          doi = {10.48550/arXiv.1004.0419},
archivePrefix = {arXiv},
       eprint = {1004.0419},
 primaryClass = {astro-ph.SR},
       adsurl = {https://ui.adsabs.harvard.edu/abs/2010ASPC..435..335R}
}

@ARTICLE{Mukai+08,
       author = {{Mukai}, K. and {Orio}, M. and {Della Valle}, M.},
        title = "{Novae as a Class of Transient X-Ray Sources}",
      journal = {\apj},
         year = 2008,
        month = apr,
       volume = {677},
       number = {2},
        pages = {1248-1252},
          doi = {10.1086/529362},
       adsurl = {https://ui.adsabs.harvard.edu/abs/2008ApJ...677.1248M}
}

@ARTICLE{Stel+25,
       author = {{Stel}, Giovanni and {Ponti}, Gabriele and {Degenaar}, Nathalie and {Sidoli}, Lara and {Mereghetti}, Sandro and {Mori}, Kaya and {Bao}, Tong and {Illiano}, Giulia and {Mondal}, Samaresh and {Reynolds}, Mark and {Jin}, Chichuan and {Lian}, Tianying and {Mandel}, Shifra and {Scaringi}, Simone and {Zhang}, Shuo and {Sanger-Johnson}, Grace and {Wijnands}, Rudy and {Miller}, Jon M. and {Kennea}, Jamie and {Zhu}, Zhenlin},
        title = "{The Very Faint X-ray Transient 4XMM J174610.7-290020 at the Galactic center}",
      journal = {arXiv e-prints},
         year = 2025,
        month = oct,
          eid = {arXiv:2510.02079},
        pages = {arXiv:2510.02079},
          doi = {10.48550/arXiv.2510.02079},
archivePrefix = {arXiv},
       eprint = {2510.02079},
 primaryClass = {astro-ph.HE},
       adsurl = {https://ui.adsabs.harvard.edu/abs/2025arXiv251002079S}
}

@ARTICLE{Mukai+17,
       author = {{Mukai}, K.},
        title = "{X-Ray Emissions from Accreting White Dwarfs: A Review}",
      journal = {\pasp},
         year = 2017,
        month = jun,
       volume = {129},
       number = {976},
        pages = {062001},
          doi = {10.1088/1538-3873/aa6736},
archivePrefix = {arXiv},
       eprint = {1703.06171},
 primaryClass = {astro-ph.HE},
       adsurl = {https://ui.adsabs.harvard.edu/abs/2017PASP..129f2001M}
}

@ARTICLE{Pastor-Marazuela+20,
       author = {{Pastor-Marazuela}, I. and {Webb}, N.~A. and {Wojtowicz}, D.~T. and {van Leeuwen}, J.},
        title = "{The EXOD search for faint transients in XMM-Newton observations: Method and discovery of four extragalactic Type I X-ray bursters}",
      journal = {\aap},
         year = 2020,
        month = aug,
       volume = {640},
          eid = {A124},
        pages = {A124},
          doi = {10.1051/0004-6361/201936869},
archivePrefix = {arXiv},
       eprint = {2005.08673},
 primaryClass = {astro-ph.HE},
       adsurl = {https://ui.adsabs.harvard.edu/abs/2020A&A...640A.124P}
}

@ARTICLE{Alizai+23,
       author = {{Alizai}, K. and {Chenevez}, J. and {Cumming}, A. and {Degenaar}, N. and {Falanga}, M. and {Galloway}, D.~K. and {in't Zand}, J.~J.~M. and {Jaisawal}, G.~K. and {Keek}, L. and {Kuulkers}, E. and {Lampe}, N. and {Schatz}, H. and {Serino}, M.},
        title = "{A catalogue of unusually long thermonuclear bursts on neutron stars}",
      journal = {\mnras},
         year = 2023,
        month = may,
       volume = {521},
       number = {3},
        pages = {3608-3624},
          doi = {10.1093/mnras/stad374},
archivePrefix = {arXiv},
       eprint = {2308.03499},
 primaryClass = {astro-ph.HE},
       adsurl = {https://ui.adsabs.harvard.edu/abs/2023MNRAS.521.3608A}
}

@ARTICLE{Williams+08,
       author = {{Williams}, Robert and {Mason}, Elena and {Della Valle}, Massimo and {Ederoclite}, Alessandro},
        title = "{Transient Heavy Element Absorption Systems in Novae: Episodic Mass Ejection from the Secondary Star}",
      journal = {\apj},
         year = 2008,
        month = sep,
       volume = {685},
       number = {1},
        pages = {451-462},
          doi = {10.1086/590056},
archivePrefix = {arXiv},
       eprint = {0805.1372},
 primaryClass = {astro-ph},
       adsurl = {https://ui.adsabs.harvard.edu/abs/2008ApJ...685..451W}
}

@ARTICLE{Page+15,
       author = {{Page}, K.~L. and {Osborne}, J.~P. and {Kuin}, N.~P.~M. and {Henze}, M. and {Walter}, F.~M. and {Beardmore}, A.~P. and {Bode}, M.~F. and {Darnley}, M.~J. and {Delgado}, L. and {Drake}, J.~J. and {Hernanz}, M. and {Mukai}, K. and {Nelson}, T. and {Ness}, J.-U. and {Schwarz}, G.~J. and {Shore}, S.~N. and {Starrfield}, S. and {Woodward}, C.~E.},
        title = "{Swift detection of the super-swift switch-on of the super-soft phase in nova V745 Sco (2014)}",
      journal = {\mnras},
         year = 2015,
        month = dec,
       volume = {454},
       number = {3},
        pages = {3108-3120},
          doi = {10.1093/mnras/stv2144},
archivePrefix = {arXiv},
       eprint = {1509.04004},
 primaryClass = {astro-ph.SR},
       adsurl = {https://ui.adsabs.harvard.edu/abs/2015MNRAS.454.3108P}
}

@ARTICLE{Page+20,
       author = {{Page}, K.~L. and {Kuin}, N.~P.~M. and {Beardmore}, A.~P. and {Walter}, F.~M. and {Osborne}, J.~P. and {Markwardt}, C.~B. and {Ness}, J.-U. and {Orio}, M. and {Sokolovsky}, K.~V.},
        title = "{The 2019 eruption of recurrent nova V3890 Sgr: observations by Swift, NICER, and SMARTS}",
      journal = {\mnras},
         year = 2020,
        month = dec,
       volume = {499},
       number = {4},
        pages = {4814-4831},
          doi = {10.1093/mnras/staa3083},
archivePrefix = {arXiv},
       eprint = {2010.01001},
 primaryClass = {astro-ph.HE},
       adsurl = {https://ui.adsabs.harvard.edu/abs/2020MNRAS.499.4814P}
}

@ARTICLE{Su+2026,
       author = {{Su}, Zhao and {Li}, Zhiyuan},
        title = "{Hydrodynamic evolution and detectability of nova remnants in the Galactic centre}",
      journal = {\mnras},
         year = 2026,
        month = feb,
       volume = {545},
       number = {4},
          eid = {staf2247},
        pages = {staf2247},
          doi = {10.1093/mnras/staf2247},
archivePrefix = {arXiv},
       eprint = {2512.16316},
 primaryClass = {astro-ph.HE},
       adsurl = {https://ui.adsabs.harvard.edu/abs/2026MNRAS.545S2247S}
}

@ARTICLE{RodriguezGil+23,
       author = {{Rodr{\'\i}guez-Gil}, Pablo and {Corral-Santana}, Jes{\'u}s M. and {El{\'\i}as-Rosa}, N. and {G{\"a}nsicke}, Boris T. and {Hernanz}, Margarita and {Sala}, Gloria},
        title = "{The orbital period of the recurrent nova V2487 Oph revealed}",
      journal = {\mnras},
         year = 2023,
        month = dec,
       volume = {526},
       number = {4},
        pages = {4961-4975},
          doi = {10.1093/mnras/stad3124},
archivePrefix = {arXiv},
       eprint = {2310.05877},
 primaryClass = {astro-ph.SR},
       adsurl = {https://ui.adsabs.harvard.edu/abs/2023MNRAS.526.4961R}
}

@ARTICLE{Ness+07,
       author = {{Ness}, J.-U. and {Starrfield}, S. and {Beardmore}, A.~P. and {Bode}, M.~F. and {Drake}, J.~J. and {Evans}, A. and {Gehrz}, R.~D. and {Goad}, M.~R. and {Gonzalez-Riestra}, R. and {Hauschildt}, P. and {Krautter}, J. and {O'Brien}, T.~J. and {Osborne}, J.~P. and {Page}, K.~L. and {Sch{\"o}nrich}, R.~A. and {Woodward}, C.~E.},
        title = "{The SSS Phase of RS Ophiuchi Observed with Chandra and XMM-Newton. I. Data and Preliminary Modeling}",
      journal = {\apj},
         year = 2007,
        month = aug,
       volume = {665},
       number = {2},
        pages = {1334-1348},
          doi = {10.1086/519676},
archivePrefix = {arXiv},
       eprint = {0705.1206},
 primaryClass = {astro-ph},
       adsurl = {https://ui.adsabs.harvard.edu/abs/2007ApJ...665.1334N}
}

@ARTICLE{Orio+23,
       author = {{Orio}, Marina and {Gendreau}, Keith and {Giese}, Morgan and {Luna}, Gerardo Juan M. and {Magdolen}, Jozef and {Strohmayer}, Tod E. and {Zhang}, Andy E. and {Altamirano}, Diego and {Dobrotka}, Andrej and {Enoto}, Teruaki and {Ferrara}, Elizabeth C. and {Ignace}, Richard and {Heinz}, Sebastian and {Markwardt}, Craig and {Nichols}, Joy S. and {Parker}, Michael L. and {Pasham}, Dheeraj R. and {Pei}, Songpeng and {Pradhan}, Pragati and {Remillard}, Ron and {Steiner}, James F. and {Tombesi}, Francesco},
        title = "{The RS Oph Outburst of 2021 Monitored in X-Rays with NICER}",
      journal = {\apj},
         year = 2023,
        month = sep,
       volume = {955},
       number = {1},
          eid = {37},
        pages = {37},
          doi = {10.3847/1538-4357/ace9bd},
archivePrefix = {arXiv},
       eprint = {2307.11485},
 primaryClass = {astro-ph.HE},
       adsurl = {https://ui.adsabs.harvard.edu/abs/2023ApJ...955...37O}
}

@ARTICLE{Schaefer+10,
       author = {{Schaefer}, Bradley E.},
        title = "{Comprehensive Photometric Histories of All Known Galactic Recurrent Novae}",
      journal = {\apjs},
         year = 2010,
        month = apr,
       volume = {187},
       number = {2},
        pages = {275-373},
          doi = {10.1088/0067-0049/187/2/275},
archivePrefix = {arXiv},
       eprint = {0912.4426},
 primaryClass = {astro-ph.SR},
       adsurl = {https://ui.adsabs.harvard.edu/abs/2010ApJS..187..275S}
}

@ARTICLE{Schaefer+09,
       author = {{Schaefer}, Bradley E.},
        title = "{Orbital Periods for Three Recurrent Novae}",
      journal = {\apj},
         year = 2009,
        month = may,
       volume = {697},
       number = {1},
        pages = {721-729},
          doi = {10.1088/0004-637X/697/1/721},
archivePrefix = {arXiv},
       eprint = {0903.1349},
 primaryClass = {astro-ph.SR},
       adsurl = {https://ui.adsabs.harvard.edu/abs/2009ApJ...697..721S}
}

@ARTICLE{Helton+08,
       author = {{Helton}, L.~A. and {Woodward}, C.~E. and {Vanlandingham}, K. and {Schwarz}, G.~J.},
        title = "{V2491 Cygni}",
      journal = {Central Bureau Electronic Telegrams},
         year = 2008,
        month = may,
       volume = {1379},
        pages = {1},
       adsurl = {https://ui.adsabs.harvard.edu/abs/2008CBET.1379....1H}
}

@ARTICLE{Page+10,
       author = {{Page}, K.~L. and {Osborne}, J.~P. and {Evans}, P.~A. and {Wynn}, G.~A. and {Beardmore}, A.~P. and {Starling}, R.~L.~C. and {Bode}, M.~F. and {Ibarra}, A. and {Kuulkers}, E. and {Ness}, J.-U. and {Schwarz}, G.~J.},
        title = "{Swift observations of the X-ray and UV evolution of V2491 Cyg (Nova Cyg 2008 No. 2)}",
      journal = {\mnras},
         year = 2010,
        month = jan,
       volume = {401},
       number = {1},
        pages = {121-130},
          doi = {10.1111/j.1365-2966.2009.15681.x},
archivePrefix = {arXiv},
       eprint = {0909.1501},
 primaryClass = {astro-ph.HE},
       adsurl = {https://ui.adsabs.harvard.edu/abs/2010MNRAS.401..121P}
}

%\begin{thebibliography}{99}
%\end{thebibliography}

%%%%%%%%%%%%%%%%%%%%%%%%%%%%%%%%%%%%%%%%%%%%%%%%%%

%%%%%%%%%%%%%%%%% APPENDICES %%%%%%%%%%%%%%%%%%%%%

\appendix

\section{Dust scattering effect}
\label{sec:dust}
Considering the large foreground hydrogen column density towards Swift J174610, a significant amount of intervening dust is expected, which can give rise to dust scattering.
The intensity of the scattered light observed at an angle $\theta$, under the single-scattering approximation \citep{Mathis+91}, is expressed as

$$I_{\rm sca}(\theta)=F_X N_{\rm H,sca}\int^{E_{\rm max}}_{E_{\rm min}}S(E)\int^1_0\frac{f(x)}{(1-x)^2}$$

$$\int^{a_{\rm max}}_{a_{\rm min}}n(a)\times \frac{{\rm d}\sigma_{\rm sca}(a,x,E,\theta)}{{\rm d}\Omega}{\rm d}a{\rm d}x{\rm d}E$$ 

where $x$ is the fractional distance of the assumed dust layer from the observer.
The corresponding scattering time delay is approximately
$$\Delta t(x,\theta,l)\approx (1.21 {~\rm s})\frac{x}{1-x}\theta({\rm arcsec})^2l({\rm kpc})$$

where $l$ is the distance to the source.

According to \citet{Jin+17}, two major scattering components lie along the line of sight between Earth and the GC, located at a fractional distance of $x=0.89$ and $x=0.36$, respectively. 
These contribute $\sim26\%$ and $\sim74\%$ of the total dust column, respectively.
For the adopted Swift/XRT photometry aperture of $30^{\prime\prime}$, we estimate that about 10\% of the observed flux arises from scattered light, and that the associated time delay is on the order of days. 
If the intrinsic light curve follows an exponential decay similar to that of RS Oph (see Sec.~\ref{sec:recurrent_nova}), the dust scattering effect would act to flatten the decay profile, because scattered photons arrive later than direct photons and thus redistribute the flux over a duration of the days. 
However, unless with some fine-tuning of the foreground dust distribution or the intrinsic light curve, such a process seems unlikely to solely produce the irregular, daily fluctuations observed during the 2024 outburst. 
This strongly suggests that the variability of Swift J174610 is intrinsic. 

\bsp	% typesetting comment
\label{lastpage}
\end{document}